\documentclass[a4paper,prb,twocolumn,superscriptaddress,notitlepage]{revtex4-1}
\usepackage[utf8]{inputenc}
\usepackage{graphicx,epsfig,color}
\usepackage{latexsym}
\usepackage{amsfonts} 
\usepackage{amsmath}
\usepackage{textcomp}
\usepackage{hyperref}
\usepackage{placeins}
\usepackage{stackrel}

\def\ket|#1>{| #1 \rangle}
\def\bra<#1|{\langle #1 |}
\def\<{\langle}
\def\>{\rangle}
\def\{{\lbrace}
\def\}{\rbrace}
\def\({\left(}
\def\){\right)}
\def\[{\left[}
\def\]{\right]}
\def\beq{\begin{equation}}
\def\eeq{\end{equation}}
\def\eff{\mathrm{eff}}

\def\{{\left\lbrace}
\def\}{\right\rbrace}

\def\eps{\varepsilon}
\def\barray{\begin{align}}
\def\earray{\end{align}}

\begin{document}

\title{Exotic correlation spread in free-fermionic states with initial
  patterns}

\author{Sudipto Singha Roy}
\affiliation{Instituto de Física Teórica UAM/CSIC, Universidad
  Autónoma de Madrid, Cantoblanco, Madrid, Spain}

\author{Giovanni Ramírez}
\affiliation{Instituto de Investigación, Escuela de Ciencias Físicas y
  Matemáticas, Universidad de San Carlos de Guatemala, Guatemala}

\author{Silvia N. Santalla}
\affiliation{Departamento de Física and Grupo Interdisciplinar de Sistemas
  Complejos (GISC), Universidad Carlos III de Madrid, Spain}

\author{Germán Sierra}
\affiliation{Instituto de Física Teórica UAM/CSIC, Universidad
  Autónoma de Madrid, Cantoblanco, Madrid, Spain}

\author{Javier Rodríguez-Laguna}
\affiliation{Departamento de Física Fundamental, UNED, Madrid, Spain}

\begin{abstract}
 We describe a relation between the light-cone velocities after a quantum quench and the internal structure of
the initial state, in the particular case of free fermions on a chain at half filling. The considered states include
short-range valence bond solids, i.e., dimerized states, and long-range states such as the rainbow. In all the
considered cases the correlations spread into one or a few well-defined light cones, each of them presenting an
effective velocity which can be read from the form factor. Interestingly, we find that the observed velocities range
from zero to the Fermi velocity and may not always be obtained from the dispersion relation for valid momenta.
\end{abstract}

\date{\today}

\maketitle
%
\section{Introduction}

The spread of correlations is one of the central issues regarding the
dynamics of quantum many-body systems. The main insight was provided
by Lieb and Robinson \cite{Lieb_Robinson.72}, when they proved
rigorously that a {\em light-cone} structure appears within the
dynamics of short-ranged Hamiltonians under some mild mathematical
conditions on the nature of the interaction. Yet, it is relevant to
ask about the effective velocity associated to the light-cone, and its
relation to the propagation velocity of {\em quasiparticles}
\cite{Calabrese.05}, which is associated to the maximal group velocity
according to the dispersion relations \cite{De_Chiara.06,Fagotti.08},
both in local and long-ranged Hamiltonians
\cite{Lauchli.08,Nezhadhaghighi.14,Buyskikh.16}, or in the case of
periodically changing Hamiltonians \cite{Eisler.08}. The time
evolution of the entanglement entropy (EE) under integrable
Hamiltonians has received special attention. For example,
non-equilibrium dynamics of EE after a sudden quantum quench has been
extensively studied for the Ising model in a transverse field
\cite{Bucciantini.14} or the XY model \cite{Fagotti.15}. Recently, the
exact time-evolution of the EE has been found for the XXZ model and
the Lieb-Liniger model, showing a velocity dependence on the
interaction parameters \cite{Alba.17,Alba.18}. The non-integrable case
presents its own challenges. For example, the Ising model subject to
both a transversal and a longitudinal field shows that the spread of
entanglement can be significatively faster than that of energy
\cite{Kim.13}. Moreover, the light-cone may fade away for some values
of the interaction parameters, related to the interpretation of the
Hamiltonian as a toy model for quark confinement \cite{Kormos.17},
without violating the Lieb-Robinson result.  It is shown that general hydrodynamical
arguments yield a natural generalization of the group velocity \cite{Perfetto.21}.  An application of the conformal field theory (CFT) framework to  quantum quench  in the XX chain is also discussed  \cite{jacopo_viti}.

As the previous examples show, the effective velocity of the
light-cone may vary with the interaction parameters and
form.  Yet,  in  some relatively recent works by  Giovannini \emph{et al.} \cite{Giovannini.15} and Bouchard \emph{et al.} \cite{Bouchard.16} it is
  shown that spatially structured light-beams may propagate in vacuum
  with a speed lower than the speed of light, due to internal
  interference effects which give rise to an effective index of
  refraction. Thus, in some cases the light-cone velocity may depend
  significatively on the nature of the initial state. For example, it
  is known that thermal states present light-cone velocities
  correlated with the excess density of energy after a quench to the
  XXZ model \cite{Bonnes.14}. Along with this, it has been shown that
  the presence of entanglement in the initial state can help in
  enhancing and accelerating the growth of correlations
  \cite{Kastner.14}. Moreover, we should stress the recent work by
  Najafi \emph{et al.} \cite{Viti.18}, where it is shown that initial
  states with a spatial periodicity can present a lower light-cone
  velocity under an XY Hamiltonian.

 In this article, we extend the previous works by
  characterizing the spread of correlations in quantum states
  presenting different types of spatial structures under a spinless
  free-fermion Hamiltonian in one-dimension (1D),  which is described in the continuum limit by a CFT. In many cases, correlations may spread into more
  than one light-cones. As we will show, the different light-cone
  velocities, which range from zero to the Fermi velocity, can be read
  from the form factor, i.e. the correlation matrix in momentum
  space. The behavior is also found to be imprinted in the growth of
  the EE where we report more than one linear stage
  of growth, with different slopes, corresponding to the passage of
  the different types of quasiparticles. We also extend the set of
  initial states, considering cases with short-range correlations,
  such as the dimerized state and a few more complex relatives, but
  also initial states with long-range correlations, such as the
  rainbow state and its variants
  \cite{Vitagliano.10,Ramirez.14,Ramirez.15,Laguna.16,
    Laguna.17,Tonni.18,Samos.19,MacCormack.19,Samos.20,Samos.21}. By
  extending the formalism in the continuum limit, we show that in all
  the cases the structure of the correlation matrix away from the
  light-cone presents universal signatures: the correlations along the
  light-cone decay as $t^{-1/3}$.

This article is organized as follows. In section \ref{sec:model} we
describe our model and initial states. Section \ref{sec:theory} leads
to our main result, showing how a spatial pattern in the initial
correlation matrix may result in an effective velocity different from
the Fermi velocity. In section \ref{sec:entropy} we discuss the
implications of our results towards the time-evolution of the 
EE of different blocks. Section \ref{sec:kpz}
discusses the universal features of the time-evolved correlation
matrix away from the light-cone. We finish the paper in
Sec. \ref{sec:conclusions} summarizing our conclusions and suggestions
for further work.


\section{Model Hamiltonian and Initial States}
\label{sec:model}

Our dynamics will be governed by the following free-fermionic
Hamiltonian on a chain of size $N$,

\beq
H= -\frac{1}{2}\sum_{i=1}^N c^\dagger_i c_{i+1}  + \text{h.c.},
\label{eq:ham}
\eeq
where $c_i$ is the fermionic annihilation operators at site $i$, and
where periodic boundaries are in effect, $c_{N+1}=c_1$. Let us define
fermionic operators $d_k$ with a well defined momentum

\begin{equation}
  d_k = \frac{1}{\sqrt N} \sum_{j=1}^N e^{-ijk} c_j,\;
  \label{eq:momentum}
\end{equation}
where $k$ ranges over the set of valid momenta,

\begin{equation}
  k =\{ \frac{2m\pi}{N} \middle| m = 0,1,2,\dots,N-1 \},
  \label{eq:valid_momenta}
\end{equation}
transforming Hamiltonian \eqref{eq:ham} into

\begin{equation}
  H = \sum_k \eps_k  \, d^\dagger_k d_k, \;
  \label{eq:ham_diagonal}
\end{equation}
with eigenvalues $\eps_k =-\cos k$.

\bigskip

\begin{figure}
  \includegraphics[width=8cm]{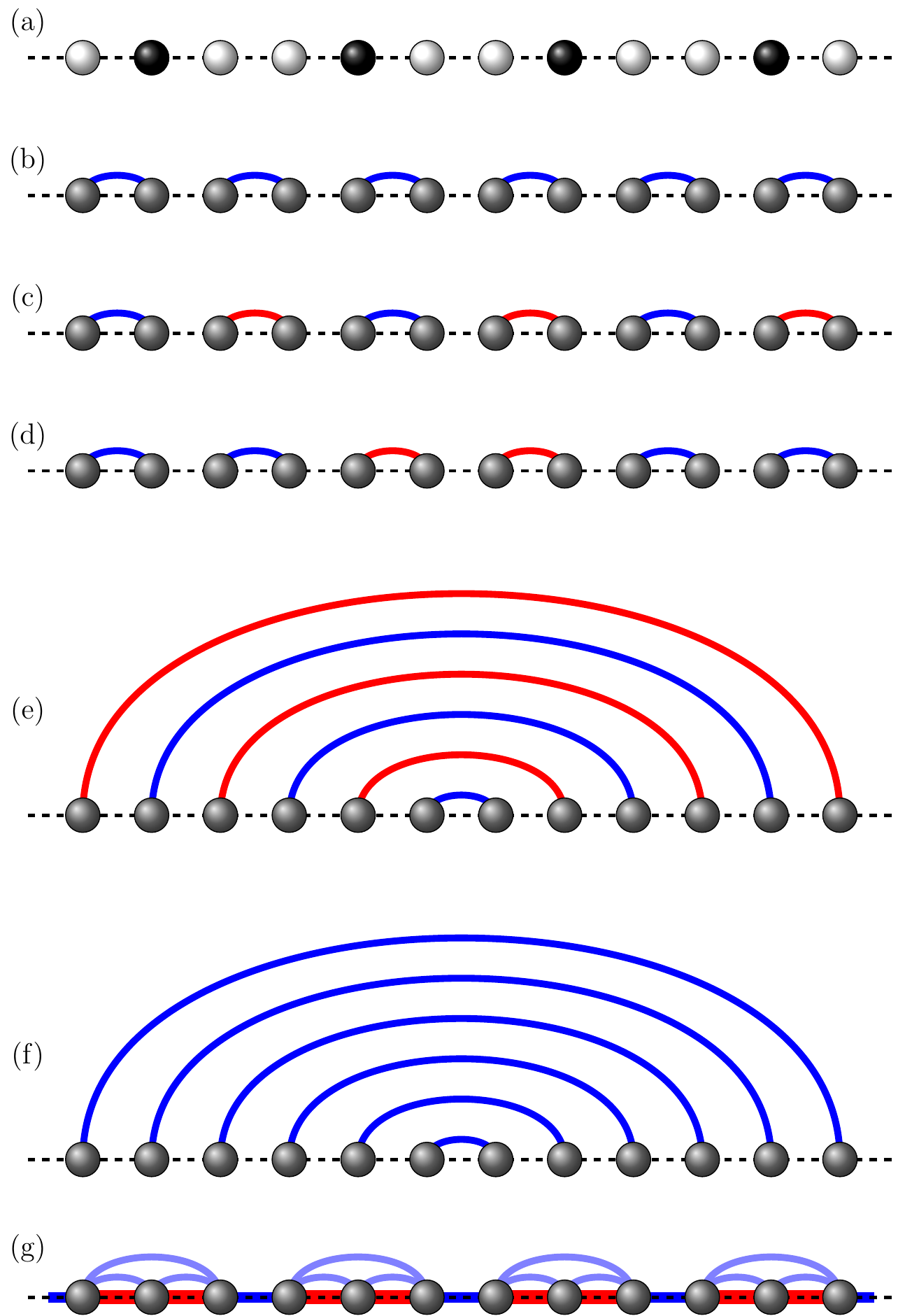}
  \caption{Illustration of (a) Wigner state with $P=3$; (b) Dimer
    state; Dimer-$q$ states, which present a spatial periodicity
    with $P=4q$ sites, for (c) $q=1$ and (d) $q=2$; (e) {\em Rainbow} state,
    whose bonds alternate signs $\eta_p$, and (f) {\em Frozen rainbow}
    state, whose bonds always have the same sign $\eta_p$; (g)
    Island-$P$ states, which present spatial periodicity with $P$, for
    $P=3$.}
  \label{fig:illust}    
\end{figure}

Let us choose a set of engineered states presenting spatial patterns,
which we will allow to evolve under the action of Hamiltonian
\eqref{eq:ham}. Our first family of initial states are the {\em Wigner
  crystals} with period $P$, illustrated in Fig. \ref{fig:illust} (a)
and defined by

\beq
\ket|W_P>=\prod_{i=1}^{N/P} c^\dagger_{Pi} \ket|0>,
\label{eq:wigner}
\eeq
where $N$ is divisible by $P$, and $1/P$ denotes the filling
fraction. In this case, the group velocity for the excitations is
given by

\beq
v_g=\left.{\partial\eps_k\over\partial k}\right|_{k_F}=\sin\({\pi\over
  P}\).
\label{eq:vlow}
\eeq
This will be our only example away from half-filling. Most of our
initial states will be {\em valence bond states} (VBS), defined by

\beq
\ket|V> \equiv 2^{-N/4} \prod_{p=1}^{N/2} (c_{l_p}^\dagger +
\eta_p \,c^\dagger_{r_p}) \ket|0>,
\label{eq:vbs}
\eeq
where the bond $p$ connects sites $l_p$ and $r_p$, with a relative
phase $\eta_p$. We will assume that for all $i\in\{1,\cdots,N\}$ there
is a unique $p$ that satisfies either $l_p=i$ or $r_p=i$, implying
that all sites are part of a unique bond. Moreover, let us define the
application $\sigma:\{1,\cdots,N\}\mapsto \{1,\cdots,N\}$ such that
$\sigma_i$ yields the index of the partner of $i$, i.e.

\beq
\sigma_i=j\; \Leftrightarrow \; \exists p\; |\; (l_p=i \wedge r_p=j) \vee
(l_p=j \wedge r_p=i).
\eeq 
%
Our first VBS example is just the {\em dimer state}, Fig. \ref{fig:illust} (b)

\beq
\ket|D>= 2^{-N/4} \prod_{p=1}^{N/2}
\(c_{2p-1}^\dagger + \,c^\dagger_{2p}\) \ket|0>,
\label{eq:dimer}
\eeq
i.e. $l_p=2p-1$, $r_p=2p$ and $\eta_p=1$ in Eq. \eqref{eq:vbs}. Of
course, $\sigma_i=i+1$ when $i$ is odd, and $i-1$ when $i$ is even. We
will also consider some interesting generalizations, such as the {\em
  dimer-$q$} state, illustrated in Fig. \ref{fig:illust} (c)-(d) and
defined by

\beq
\ket|D_q>= 2^{-N/4} \prod_{p=1}^{N/2} \(c^\dagger_{2p-1} + \Theta(p,q)\,
c_{2p}^\dagger\) \ket|0>,
\label{eq:dimerq}
\eeq
where $\Theta(p,q)=(-1)^{\lfloor p/q\rfloor \mod 2}$, i.e.: it
alternates $q$ bonds with $\eta_p=+1$ sign and $q$ bonds with
$\eta_p=-1$.  Therefore, the pattern repeats itself after exactly $P=4q$
sites. Next, we consider the {\em rainbow state}
\cite{Vitagliano.10,Ramirez.14,Ramirez.15,Laguna.16,Laguna.17,Tonni.18,Samos.19,MacCormack.19,Samos.20},
which is formed by a concentric set of bonds and presents maximal
entropy between its left and right halves, as it is shown in
Fig. \ref{fig:illust} (e). It is defined as

\beq
\ket|R>= 2^{-N/4} \prod_{i=1}^{N/2}
\(c_i^\dagger+(-1)^{N/2+i}\,c^\dagger_{N+1-i}\) \ket|0>.
\label{eq:rainbow}
\eeq
The rainbow state has received a great deal of attention because it
can be built as the ground state (GS)  of a deformed local Hamiltonian in
the limit in which the inhomogeneity is large. We should stress that
the form \eqref{eq:rainbow} describes the GS of some spin chains, such
as the XX, XXZ or Ising chains \cite{Laguna.16,Samos.19,Samos.21},
after a Jordan-Wigner (JW) transformation has been applied, as it can
be shown making use of the strong disorder renormalization group
(SDRG) devised by Dasgupta and Ma \cite{Dasgupta.80}. The alternating
character of the signs of its bonds can be understood in terms of the
non-local nature of the JW transformation. Of course, it makes sense
to define a rainbow state {\em without} sign alternation, which we
will call the {\em frozen rainbow} for reasons to be understood later,
see Fig. \ref{fig:illust} (f).

Our last state will not be a VBS, yet it presents an interesting
spatial periodic pattern, as shown in Fig. \ref{fig:illust} (g). We
will call it an {\em island-$P$} state, $\ket|I_P>$, and it is the GS
of a Hamiltonian that can be written by weakening every $P$-th hopping
amplitude from our original Hamiltonian \eqref{eq:ham},

\beq
H_P=H+{\gamma\over 2}\sum_{i=1}^{N/P} \(c^\dagger_{Pi}c_{Pi+1} + \text{h.c.}\),
\label{eq:island_ham}
\eeq
for $\gamma\to 1^-$ (but $\gamma\neq 1$ to avoid degeneracy), and $N$
a multiple of $P$.

\bigskip

We would like to stress that most of these states are invariant under
a spatial translation of $P$ sites, but not the rainbow
states. Moreover, all of them can be described as Slater determinants,
or Gaussian states, and as such they can be fully characterized by
their correlation matrix, via Wick's theorem. In the case of a VBS
defined by Eq. \eqref{eq:vbs} we have

\beq
C_{j,j'} = \<V|c^\dagger_j c_{j'}|V\> =
\frac{1}{2} \( \delta_{j,j'} + \eta_{p(j)} \delta_{j,\sigma(j')} \).
\label{eq:corr_vbs}
\eeq
Let us notice that all the considered states at half-filling (i.e. all
of them except the Wigner states with $P>2$) are GS of Hamiltonians with
particle-hole symmetry, which implies that their density is exactly
$\<c^\dagger_i c_i\>=1/2$.


\section{Spatial patterns and light-cone velocities}
\label{sec:theory}

In this section we provide the main result of this work, establishing
a link between the spatial pattern of the initial state and the
light-cone velocity or velocities needed to describe the time-evolved
correlation matrix.

\subsection{Free fermion dynamics}

Let us describe the necessary set-up to analyze the time dynamics of
the initial states introduced in the previous section under the
free-fermionic Hamiltonian defined in Eq. \eqref{eq:ham}. Let us
consider an initial state $\ket|\psi>$, with correlation matrix
$C_{j,j'}$. After a time $t$, we will have

\beq
\ket|\psi(t)> =e^{-itH} \ket|\psi>.
\eeq
Since all the considered states are Gaussian, we may characterize the
time evolution from the two-point correlator,

\beq
C_{j,j'}(t) = \bra<\psi(t)|c^\dagger_j c_{j'}\ket|\psi(t)>
= \bra<\psi(0)| c^\dagger_j(t) c_{j'}(t) \ket|\psi(0)>,
\label{eq:evol_corr1}
\eeq
where $c_j(t)$ is the fermion operator in the Heisenberg picture

\begin{eqnarray}
  c^{\dagger}_j(t) &=& e^{itH} c^{\dagger}_j  e^{-itH} 
  = \frac{1}{\sqrt{N}} \sum_k e^{-ijk} e^{it\eps_k} d^{\dagger}_k, \nonumber\\
  &=& \frac{1}{N} \sum_{k,\ell} e^{-i(j-\ell)k} e^{it\eps_k}c^{\dagger}_{\ell}.
  \label{eq:evol_c}
\end{eqnarray}
Plugging this equation into Eq. \eqref{eq:evol_corr1} yields

\begin{equation}
  C_{j,j'}(t) =\frac{1}{N^2}\Big[ \sum_{\substack{k,k'\\\ell,\ell'}} 
  e^{-i(j-\ell)k +i(j'-\ell')k'}\; 
  e^{it(\eps_k -\eps_{k'})}\, C_{\ell,\ell'}\Big].
  \label{eq:evol_corr2}
\end{equation}
Let us remind the reader that particle-hole symmetry implies that
$C_{j,j}(0)=1/2$, for all our initial states at half-filling. The
density can be proved to remain constant for all time,
$C_{j,j}(t)=1/2$.


\subsection{The Dimer State}

The dimer case is specially simple and well-known, and deserves to be
carried out in some detail. Applying Eq. \eqref{eq:evol_corr2} we
obtain

\begin{eqnarray}
  C_{j,j'}(t) &=&\frac{\delta_{j,j'}}{2}+\frac{1}{4}
  \( \delta_{j,j'-1}+\delta_{j,j'+1} \) \nonumber\\&+& \frac{(-1)^{j'}}{4N}
  \sum_k \Big( e^{i(-j+j'-1)k} -e^{i(-j+j'+1) k} \Big) \nonumber\\ &&e^{-2it\cos k},
  \label{eq:evol_corr_dimer}
\end{eqnarray}
which satisfies the initial condition

\begin{equation}
  C_{j,j'}(t=0) = \frac{1}{2} \( \delta_{j,j'} + \delta_{j,\sigma(j')}\),
  \label{eq:corr_dimer_initial}
\end{equation} 
in agreement with equation \eqref{eq:corr_vbs} for the dimer
state. Observe that memory of the initial state is never really lost,
because the long term time-average of the correlator yields the
original value 

\beq
\overline{C_{j,j'}}=\frac{1}{T}\int_0^T C_{j,j'}(t) dt \to
\frac{\delta_{|j-j'|,1}}{4}.
\eeq
Notice also the time dependence $e^{2it\eps_k}$ in equation
\eqref{eq:evol_corr_dimer}, which follows from the relation
$\eps_{k+\pi}=-\eps_k$. In addition to this, the term $(-1)^{j'}$ is
responsible for a parity oscillation with respect to $j'$.

\bigskip

\begin{figure}[t]
\includegraphics[width=9cm]{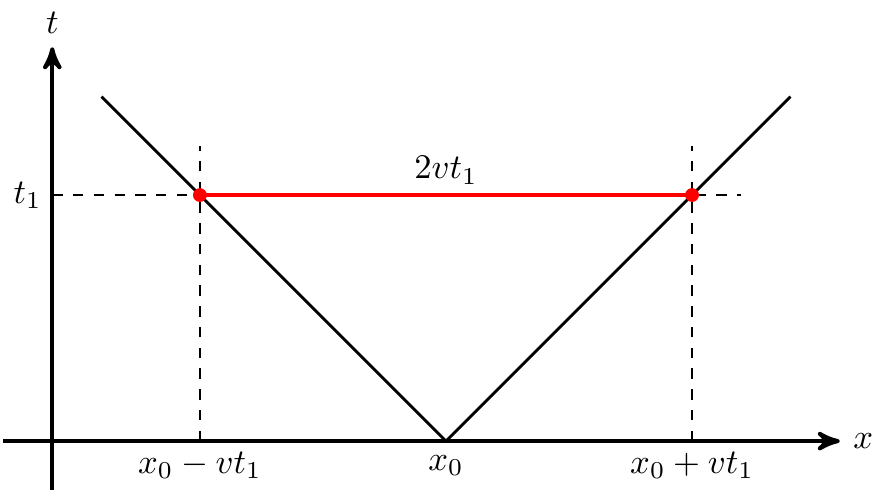}
\caption{Schematic representation of the propagation of a pair of
  quasiparticles with opposite velocities, $\pm v$, stemming from a
  point $x_0$. After a time $t_1$, the maximally correlated sites are
  located at a distance $2vt_1=v_\eff t$.}
\label{fig:schematic1}
\end{figure}

Let us consider  $N\gg 1$ limit in Eq. \eqref{eq:evol_corr_dimer},
approximating the sum by an an integral,

\beq
\frac{1}{N} \sum_{k} \rightarrow \frac{1}{2\pi} \int_{0}^{2\pi} dk,
\label{eq:continuum}
\eeq
and defining $x=j'-j$. Then, for $x>1$, we can replace
Eq. \eqref{eq:evol_corr_dimer} by

\beq
C(x,t) \simeq \frac{(-1)^{j'}}{8\pi}
   \int_{0}^{2\pi} \[ e^{i(x-1)k} -e^{i(x+1) k} \] e^{-2it\cos k}dk. 
\label{eq:cont_dimer_2}
\eeq
Now, comparing the above integrals with the standard form of the
Bessel function of first kind \cite{Abramowitz},

\beq
J_n(\nu)=\frac{e^{i\frac{n\pi}{2}}}{2\pi}
\int_0^{2\pi} e^{in\tau-i\nu \cos \tau} d\tau,
\label{eqn:Bessel}
\eeq
we get

\beq
C(x,t) \simeq 
\frac{e^{i\frac{{(j'+j+1)}\pi}{2}}}{4} \Big[J_{x-1}(2t)+J_{x+1}(2t)\Big].
\label{eq:corr_bessel}
\eeq
Notice that when the integrals are expressed in terms of Bessel
functions, the phase between the two terms changes. At $x\simeq2t$ we
can further approximate the above equation as follows,

\beq
C(x,t) \simeq 
\frac{e^{i\frac{{(j'+j+1)}\pi}{2}}}{2} J_x(2t).
\label{eq:corr_bessel_new}
\eeq
This expression shows that the correlation presents a {\em light-cone
  structure}, associated with an effective velocity $v_\eff=2$, since
$J_x(v_\eff t)\approx 0$ whenever $x\gg v_\eff t$. Notice that
$v_\eff$ is twice the Fermi velocity and apparently exceeds the
Lieb-Robinson bound. The reason can be understood through the
illustration of Fig. \ref{fig:schematic1}. Indeed, the initial state
can be thought of as a source of quasiparticle excitations which
emerge anywhere in the lattice ($x_0$) and propagate in opposite
direction with the same velocity, $\pm v$. This results in a maximal
correlation between sites located at a distance $x=2vt=v_\eff t$ from
each other.

The asymptotics of the Bessel function provides very valuable
information about the structure of the correlation functions, both
along and away from the light-cone, as we will consider in
Sec. \ref{sec:kpz}.


\begin{figure*}
  \includegraphics[width=5.8cm]{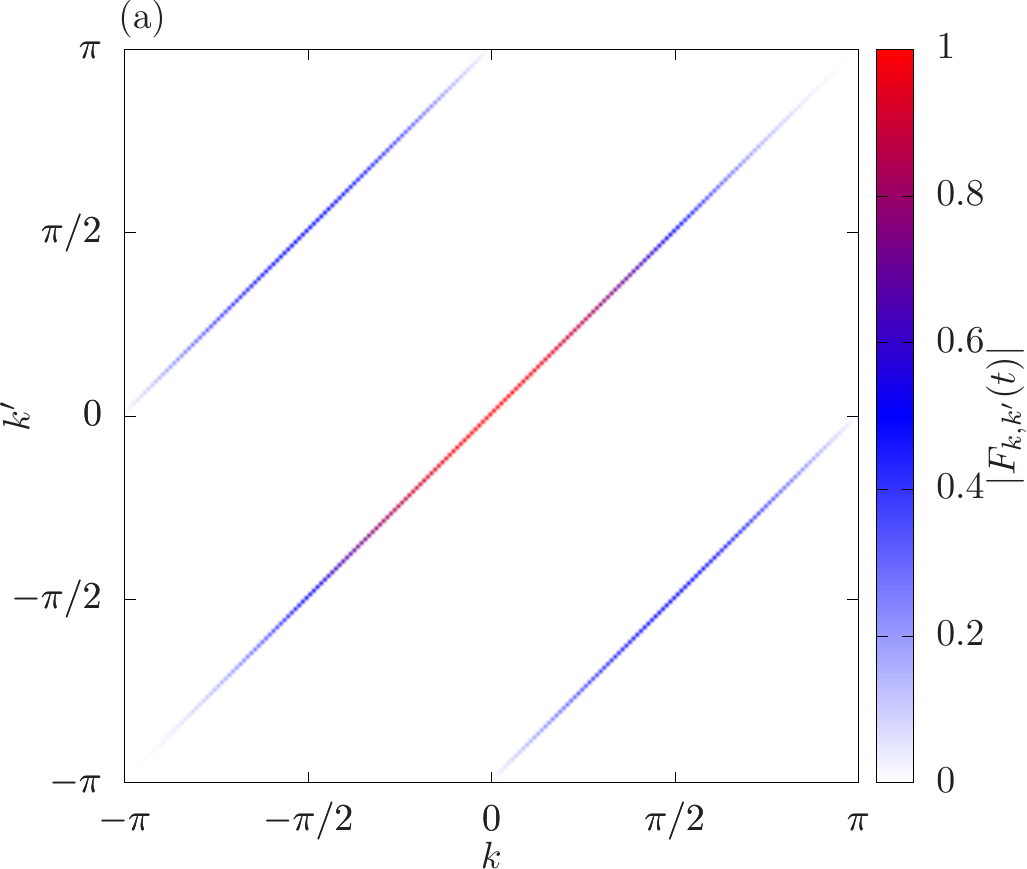}
  \includegraphics[width=5.8cm]{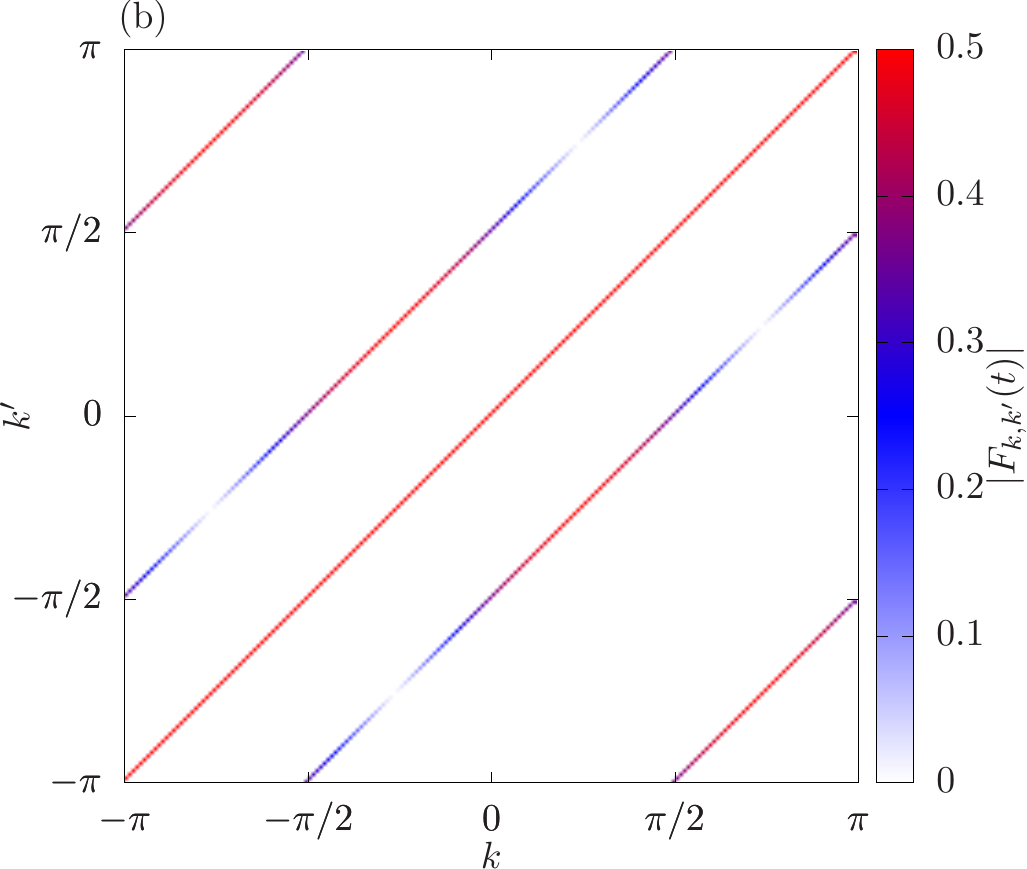}
  \includegraphics[width=5.8cm]{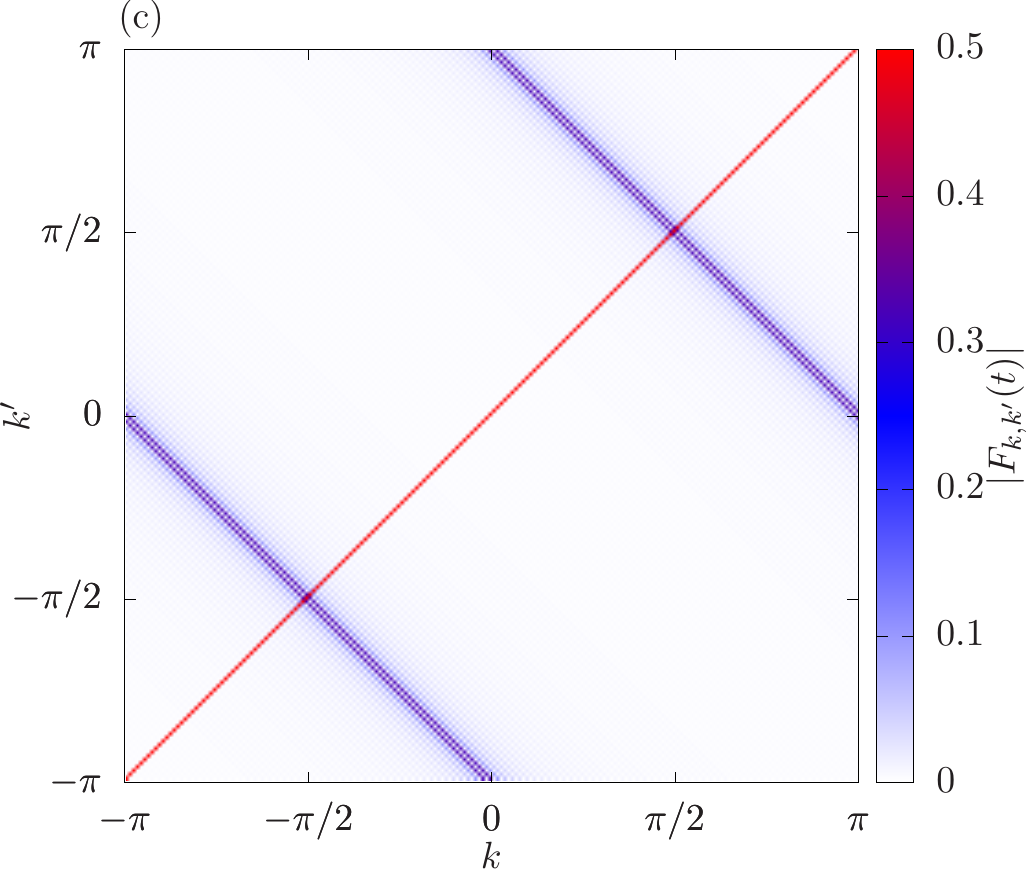}
  \includegraphics[width=5.8cm]{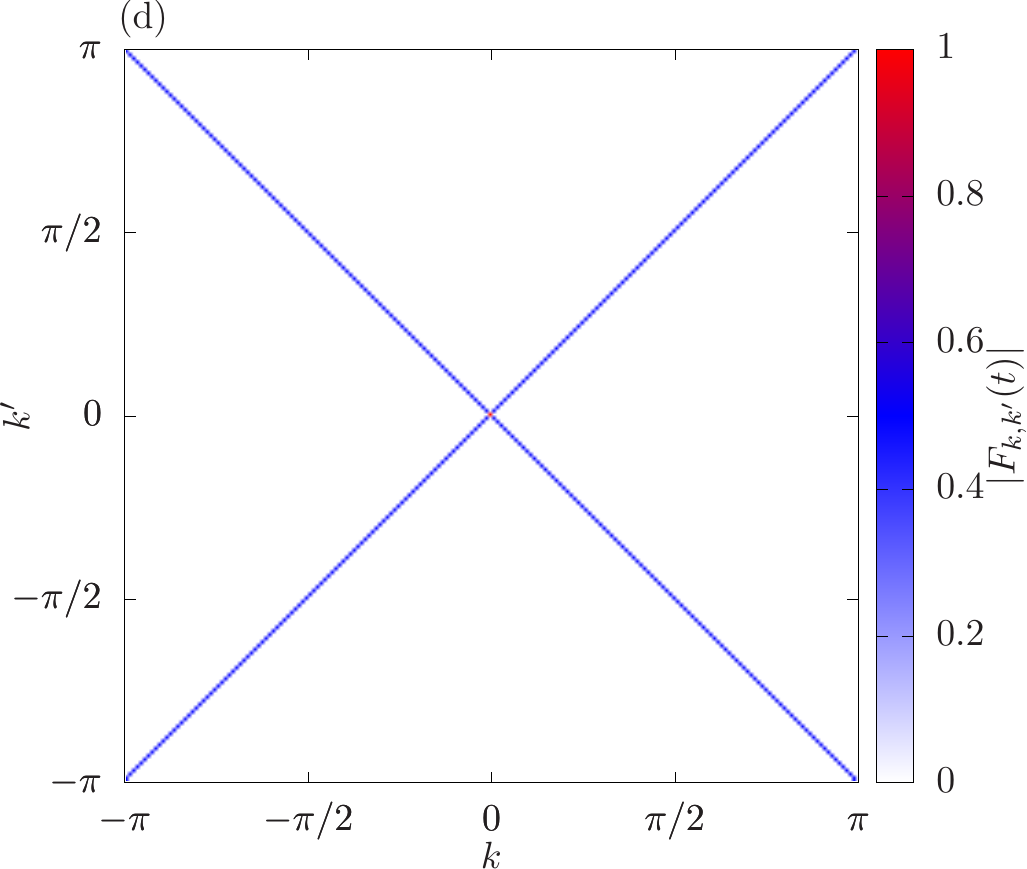}
  \includegraphics[width=5.8cm]{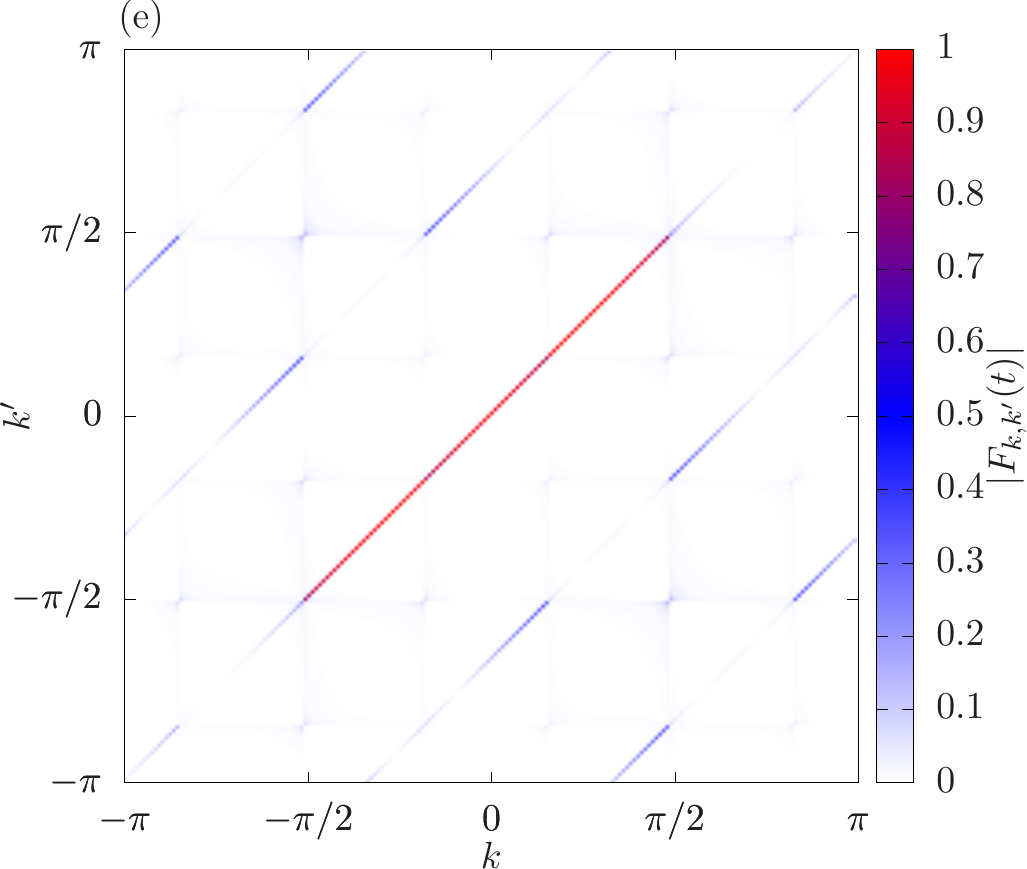}
  \caption{Absolute value of the form factors $|F_{k,k'}|$ for several
    relevant states, using $N=240$. The color code is chosen so as to
    provide a higher contrast. (a) Dimerized state,
    Eq. \eqref{eq:dimer}, (b) Dimer-1  state, Eq. \eqref{eq:dimerq},
    (c) Rainbow state, Eq. \eqref{eq:rainbow}, (d) Frozen rainbow
    state, (e) Island-3 state, using $\gamma=1-10^{-3}$.}
\label{fig:ff}
\end{figure*}

\subsection{Form Factors}

Let us define a form factor $F_{k,k'}$ associated to the initial
state $\ket|\psi(0)>$ as

\beq
F_{k,k'}\;\equiv\;\<\psi(0)|d^\dagger_k d_{k'}|\psi(0)\>\;=\;
{1\over N} \sum_{\ell,\ell'} e^{i(\ell k-\ell' k')} C_{\ell,\ell'},
\label{eq:formfactor}
\eeq
which is just the Fourier transform of the initial correlation matrix,
i.e. it corresponds to the correlation matrix in momentum
space. Notice that $|\<\psi(t)|d^\dagger_k d_{k'}|\psi(t)\>|$ is
preserved for all $k$ and $k'$ along the time evolution. Nonetheless,
our definition Eq. \eqref{eq:formfactor} only makes reference to the
initial state and does not have an absolute value. The form factor
allows to simplify the expression for the time-evolved correlation
matrix,

\beq
C_{j,j'}(t)={1\over N} \sum_{k,k'} e^{-i(jk-j'k')+it(\eps_k-\eps_{k'})} F_{k,k'}.
\label{eq:corr_ff}
\eeq
When the initial state is a VBS we can plug Eq. \eqref{eq:corr_vbs}
into Eq. \eqref{eq:formfactor} to obtain

\beq
F_{k,k'}={1\over 2}
\delta_{k,k'} + {1\over 2N} \sum_{\ell=1}^N \eta_{p(\ell)}
\(e^{i(k\ell-k'\sigma(\ell))}\).
\eeq
Let us evaluate the form factor of the states described in the
previous section. For the dimer state, Eq. \eqref{eq:dimer}, we have

\begin{align}
  F_{k,k'} =& {\delta_{k,k'}\over 2} + {1\over 2N} \sum_{p=1}^{N/2}
  \( e^{i((2p-1)k-2pk')} + e^{i(2pk-(2p-1)k')} \) \nonumber\\
  = &{\delta_{k,k'}\over 2} +{e^{ik'}+e^{-ik} \over 2N} \sum_{p=1}^{N/2}
  e^{i2p(k-k')}, \nonumber\\
  =&{\delta_{k,k'}\over 2} +
  {e^{ik'}+e^{-ik}\over 4}\(\delta_{|k-k'|=0}+\delta_{|k-k'|=\pi}\),
\end{align}
which is plotted in Fig. \ref{fig:ff} (a), where we can observe two
modulated straight lines: $k'=k$ and $k'=k\pm \pi$. In a similar way,
we can show that the form factor for the Wigner states of period $P$,
Eq. \eqref{eq:wigner} are given by

\begin{figure*}
\includegraphics[width=6.5cm]{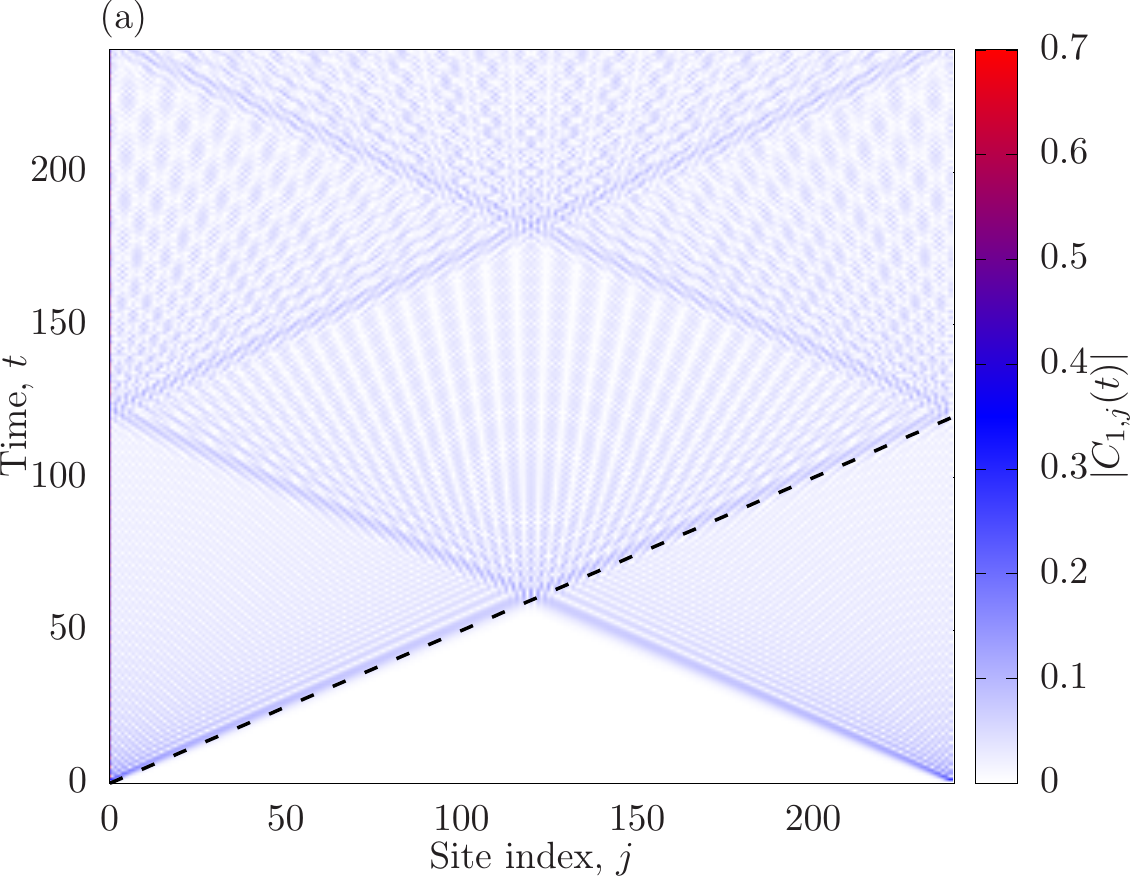}
\includegraphics[width=6.5cm]{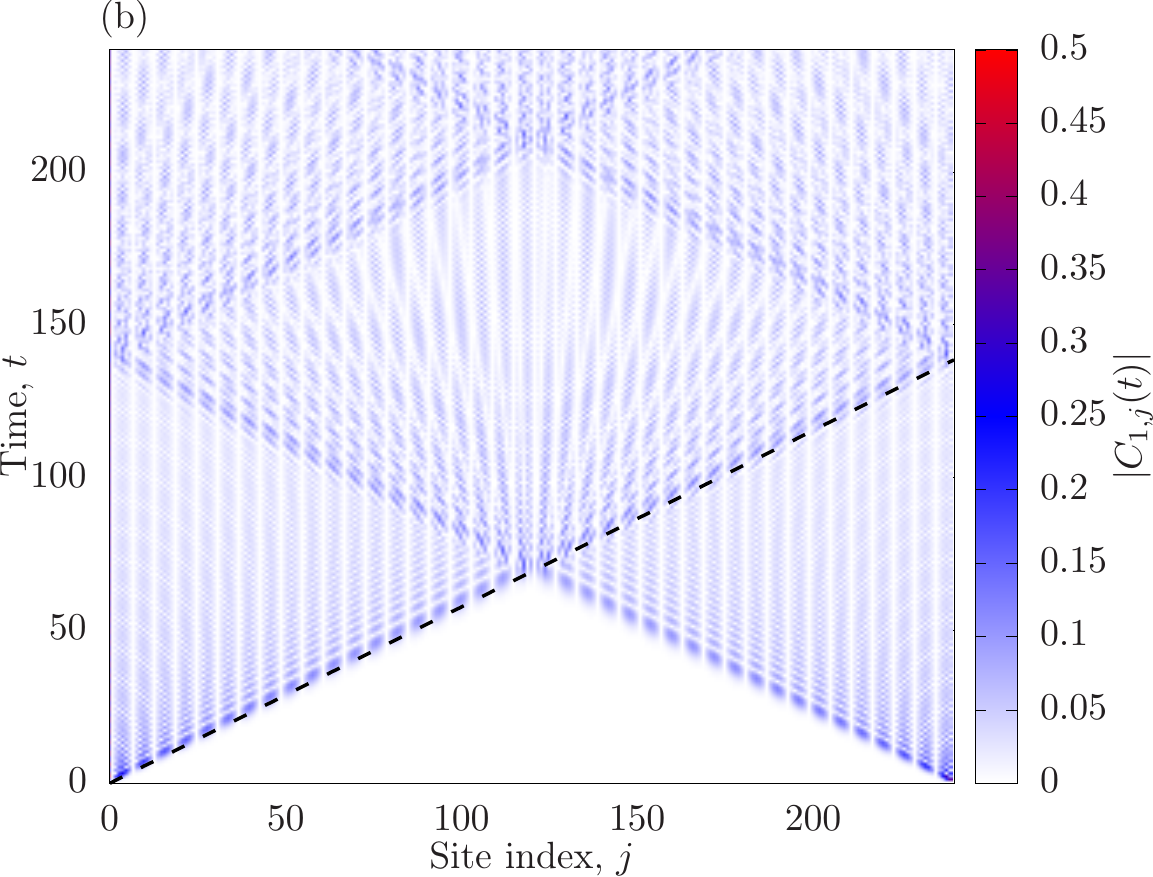}
\includegraphics[width=6.5cm]{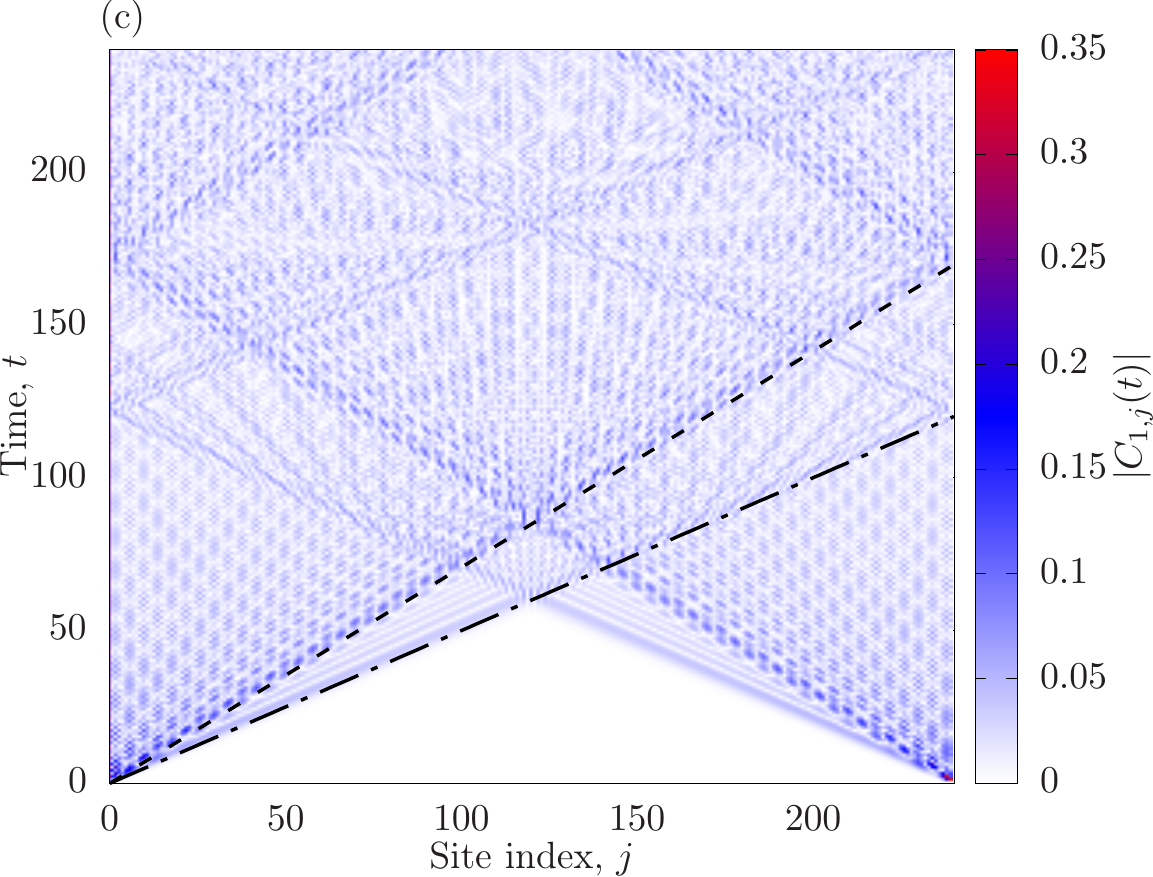}
\includegraphics[width=6.5cm]{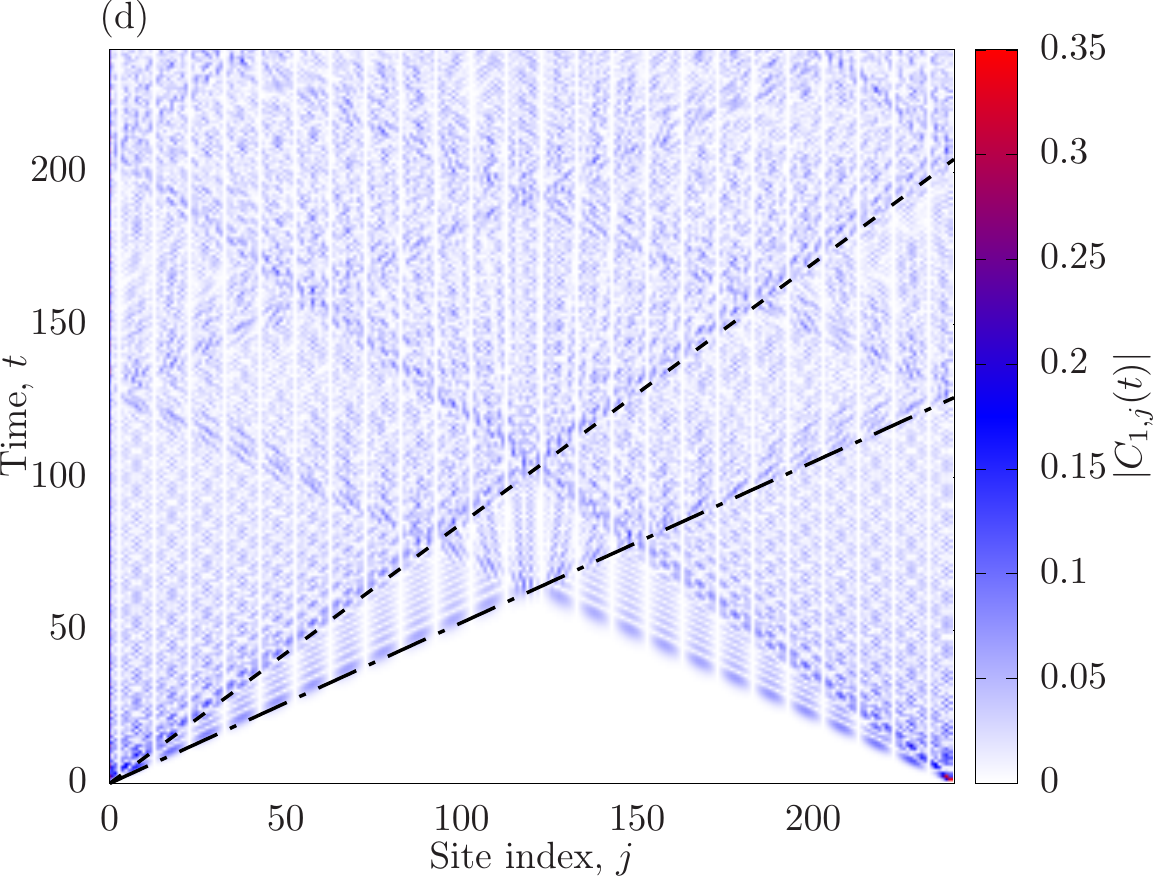}
\caption{Correlation maps $|C_{1,j}(t)|$ for Wigner crystals, for
  different values of the periodicity with $N=240$ for (a) $P=2$, (b)
  $P=3$, (c) $P=4$ and (d) $P=5$. The straight lines correspond to the
  theoretical predictions.}
\label{fig:corrmap_wigner}
\end{figure*}

\beq
F_{k,k'}=e^{-i(k-k)}{1\over 2P}\sum_{m=0}^{P-1} \delta_{|k-k'|,2\pi m/P}.
\label{eq:ff_wigner}
\eeq
The exact calculation for the other relevant states is provided in the
Appendix \ref{sec:appendix_ff}, and here we will only report the
results. For the dimer-$q$ state we obtain 

\begin{eqnarray}
&&F_{k,k'} ={\delta_{k,k'}\over 2}+{1\over 4q}\(\sum_{p=1}^q
\(\delta_{|k-k'|,{\pi(2p-1)\over 2q}}
  + \delta_{|k-k'|,2\pi-{\pi(2p-1)\over 2q}}\)\) \nonumber \\ &&\(\sum_{p=1}^q
\(e^{-i((2p-1)k-2(p-1)k')}+e^{-i(2(p-1)k-(2p-1)k')}\)\),\nonumber\\
\end{eqnarray}

which means that it presents $2q$ parallel lines of the form $k'=k\pm
(2p-1)\pi/2q$, as it is shown in Fig. \ref{fig:ff} (b). On the other
\beq
  F_{k,k'} = \frac{\delta_{k,k'}}{2}+\frac{(-1)^{N/2+1} e^{i\frac{k-k'}{2}}}{4N} 
  \frac{\(e^{i\frac{kN}{2}}+(-1)^{N/2+1}e^{-i\frac{k'N}{2}}\)^2}{\cos(\frac{k+k'}{2})}
  ,
  \label{eq:ff_rainbow}
\eeq
and it can be visualized in Fig. \ref{fig:ff} (c). The denominator
$\cos((k+k')/2)$ shows that $F_{k,k'}$ diverges whenever
$k+k'=\pm\pi$, which yields the two orthogonal lines. The frozen
rainbow has a simpler form factor, 
\beq
F_{k,k'}=\frac{\delta_{k,k'}}{2}+\frac{e^{-i(N+1)k'}\delta_{|k+k'|=0}}{2},
\eeq
which yields the two orthogonal lines, $k'=k$ and $k'=-k$, as we can
see in Fig. \ref{fig:ff} (d). Finally, we have numerically evaluated
the form factor for the island-3 state, and check that it is
approximately concentrated along straight lines of the form $k'=k\pm
2\pi/3$ and $k'=k\pm 4\pi/3$.

\begin{figure*}
\includegraphics[width=6.5cm]{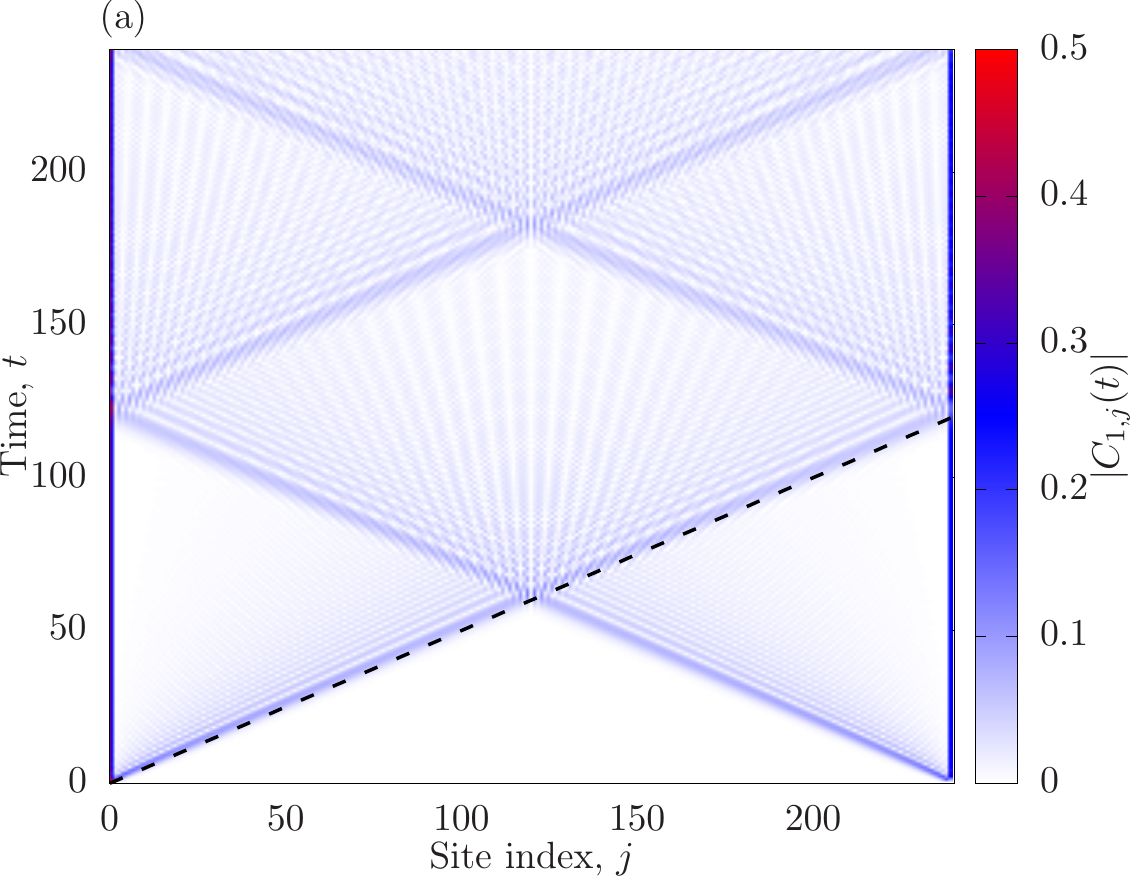}
\includegraphics[width=6.5cm]{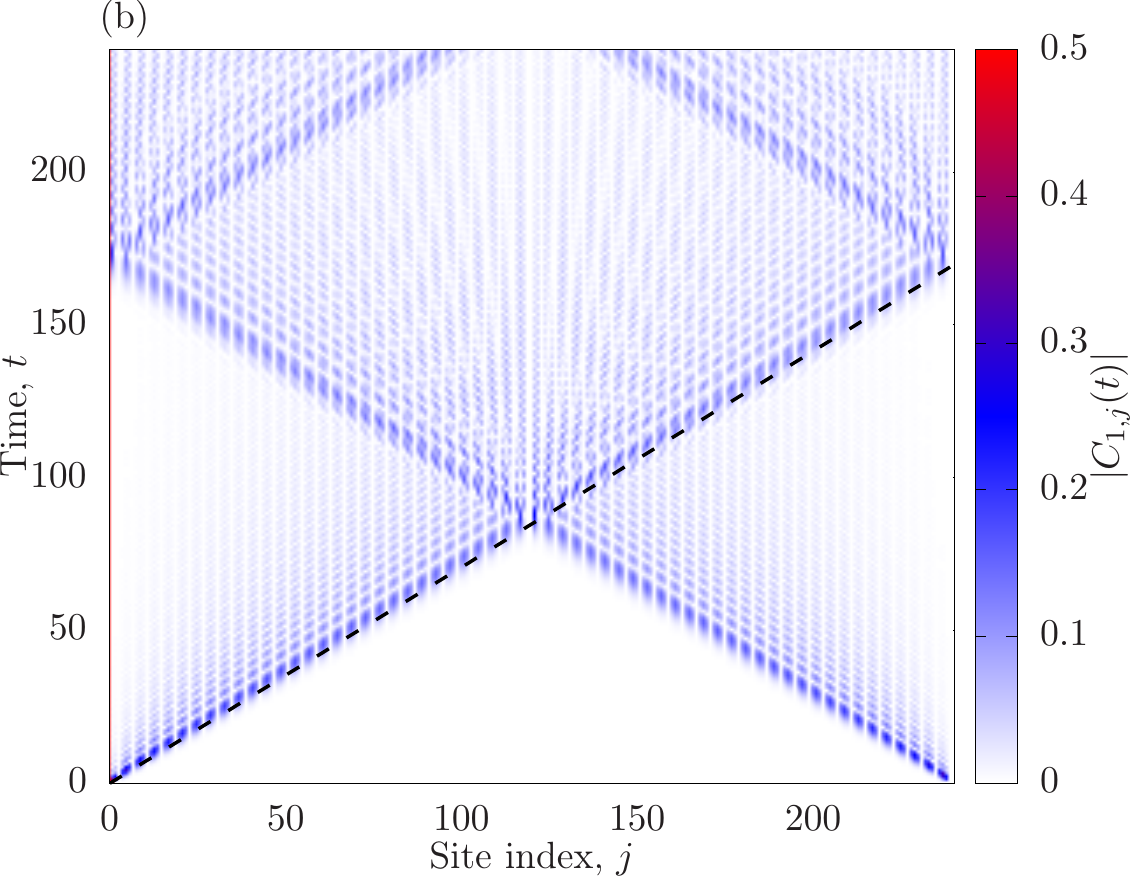}
\includegraphics[width=6.5cm]{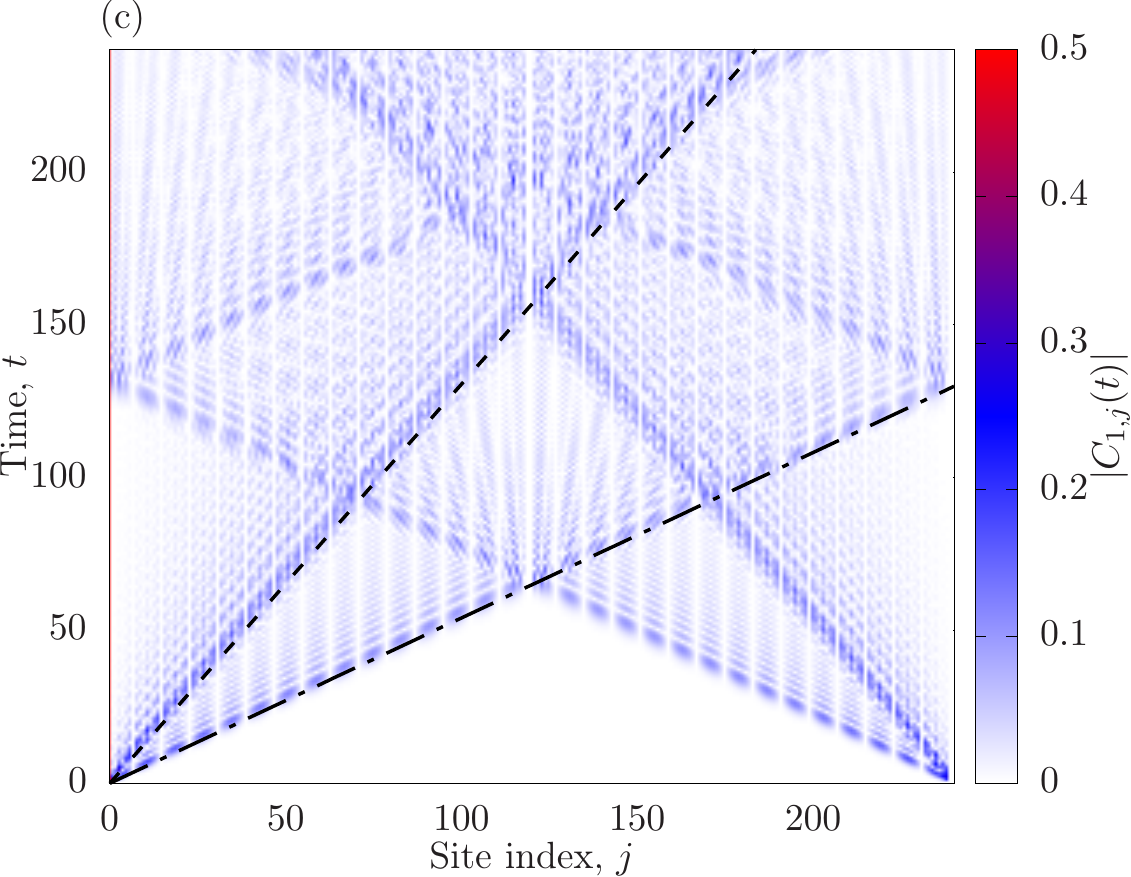}
\includegraphics[width=6.5cm]{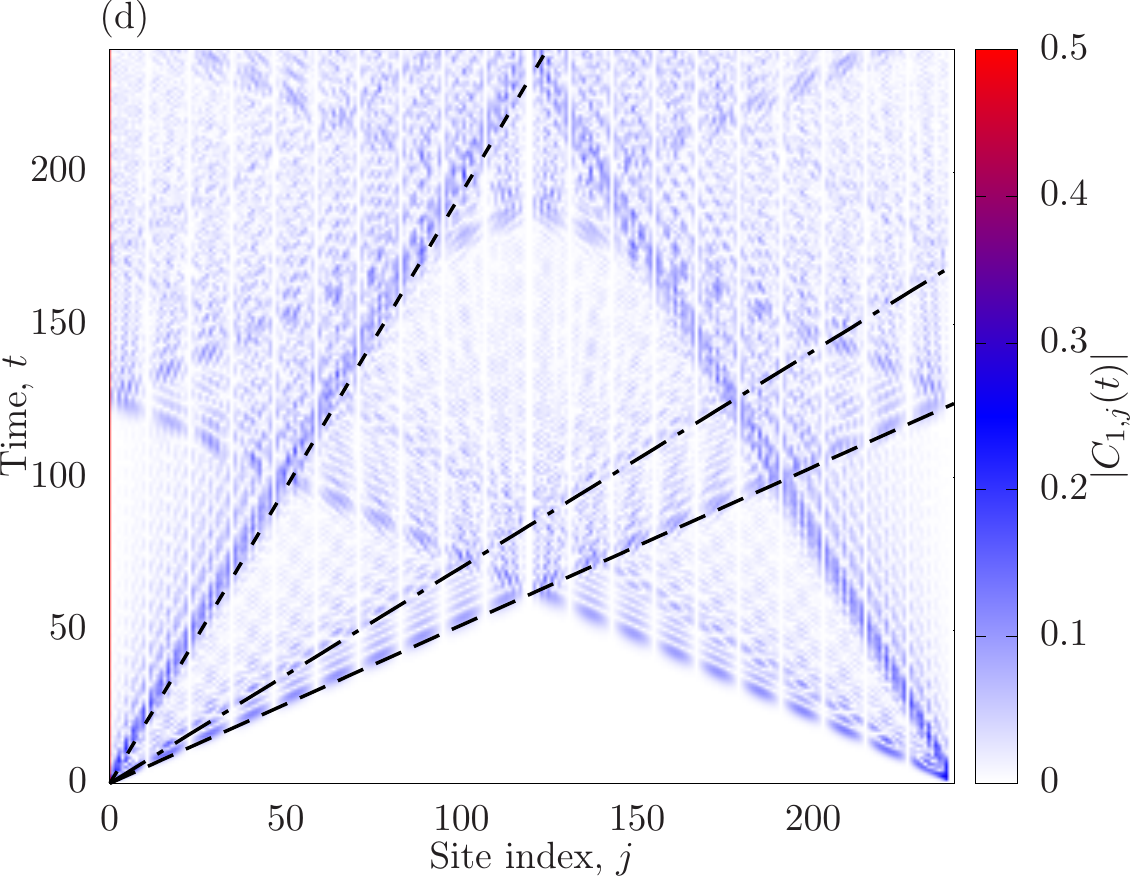}
\caption{Correlation maps $|C_{1,j}(t)|$  for dimerized states, using
  $N=240$. (a) Dimer state, (b) Dimer-1, (c) Dimer-2, and (d)
  Dimer-3. The straight lines correspond to the theoretical
  predictions.}
\label{fig:corrmap_dimer}
\end{figure*} 


\subsection{Emergence of an effective velocity}

In the previous section we realized that in many cases we can express
the form factor in the following form

\beq
F_{k,k'}\approx \sum_p F_p(k) \delta(k\pm k'+\alpha_p),
\label{eq:functional_ff}
\eeq
where $F_p(k)$ is the modulation function and $\alpha_p$ is the {\em
  phase shift}. Notice that this expression is exact in most cases,
being approximate only for the rainbow and the island states. In this
case, the time-evolved correlation matrix can be decomposed into a sum
of terms, each of which provides a light-cone with a different
effective velocity, as we will prove.

The time-evolved correlation matrix, Eq. \eqref{eq:corr_ff}, can be
written now as

\begin{eqnarray}
  C_{j,j'}(t)&\equiv & 
  \sum_p \sum_k F_p(k) e^{-i(kj\mp(k+\alpha_p)j')} e^{i(\eps_k-\eps_{\pm(k+\alpha_p)})t},\nonumber\\
  &=&\sum_p C^{(p)}_{j,j'}(t) 
\end{eqnarray}
and each $p$ term can be evaluated as

\begin{align}
C^{(p)}_{j,j'}(t) =& e^{\pm i\alpha_p j'} \sum_k F_p(k) e^{-ik(j\mp j')}
e^{-i(\cos(k)-\cos(k+\alpha_p))t},\nonumber\\
  =&e^{\pm i\alpha_p j'} \sum_k F_p(k) e^{-ik(j\mp j')}
  e^{-i2\sin\(k+\alpha_p/2\)\sin(\alpha_p/2)t},\nonumber\\
  =&e^{\pm i\alpha_p j'} \sum_q F_p(q) e^{i-((\pi-\alpha_p)/2-q)(j\mp j')}\nonumber\\&
  e^{-i\cos(q)\(2\sin(\alpha_p/2)t\)},
\end{align}
where in the last step we have defined $q=(\pi-\alpha_p)/2-k$. Notice
that the time dependence is completely encoded in the last term, and
we can define an effective velocity

\beq
v_{\eff,p}=2\sin(\alpha_p/2),
\label{eq:veff}
\eeq

thus allowing us to postulate that each straight line in the form
factor diagram yields a term in the time-evolved correlation matrix,
where the main difference is provided by the effective velocity. We
are thus led to claim that our states may present different types of
quasiparticles, characterized by different spreading velocities.

Moreover, we observe that once the velocity has been changed, the
results are quite similar to those found for the dimer case,
Eq. \eqref{eq:corr_bessel_new}, thus allowing us to conjecture that
the structure of the correlation functions will be similar in all the
considered cases, once the time axis is scaled appropriately.

\bigskip

Let us check numerically the validity of expression \eqref{eq:veff},
evaluating the time-evolution of the correlation matrix of the states
discussed in Sec. \ref{sec:model}. In all the cases we will show the
correlation $|C_{1,j}(t)|$ using a colormap, with the second index in
the horizontal axis and time in the vertical one.

Let us start with the Wigner crystals of period $P$, given in
Eq. \eqref{eq:wigner}, even though they are not at half-filling for
$P>2$. The system size has been chosen in all the cases to be a
multiple of $P$. Our theoretical prediction in this case is very
clear, because the linear structure in the form factor is exact: the
correlation matrix contains several terms, one corresponding to each
line. The velocities are always given by

\begin{figure*}
\includegraphics[width=5.8cm]{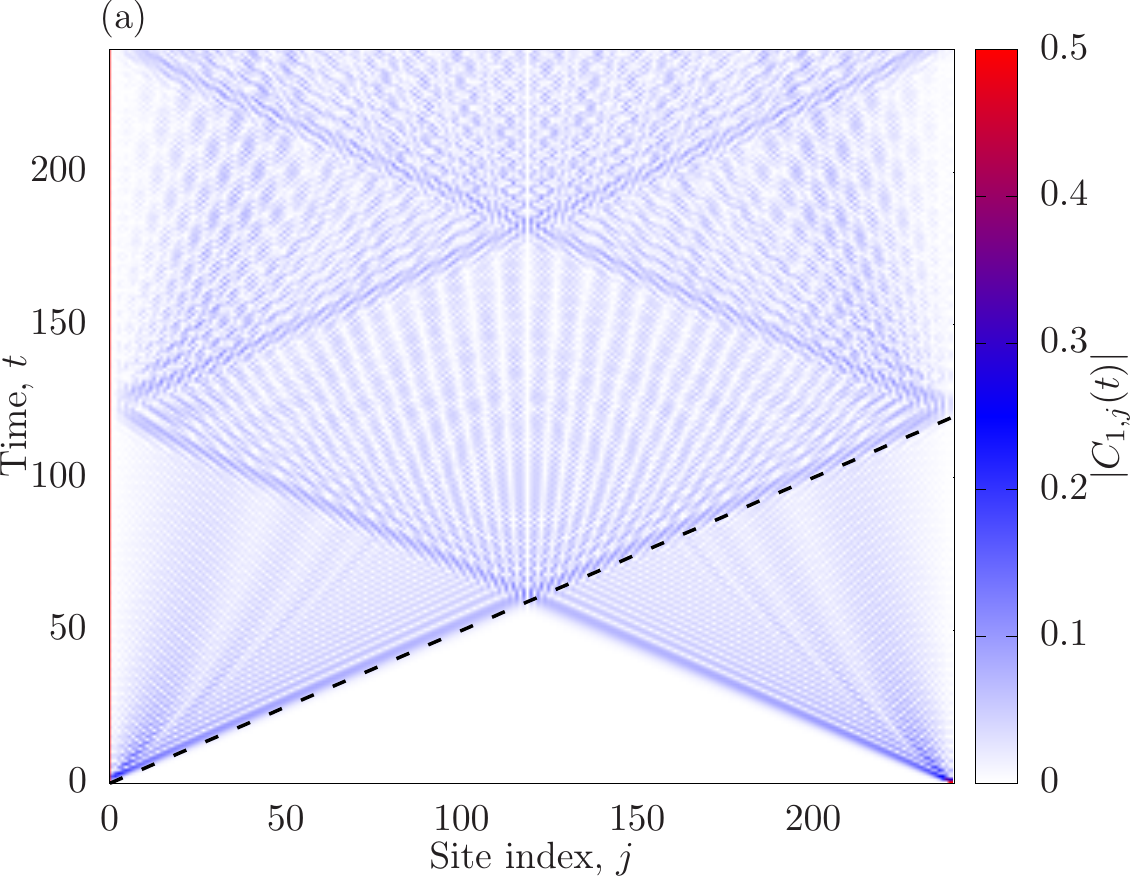}
\includegraphics[width=5.8cm]{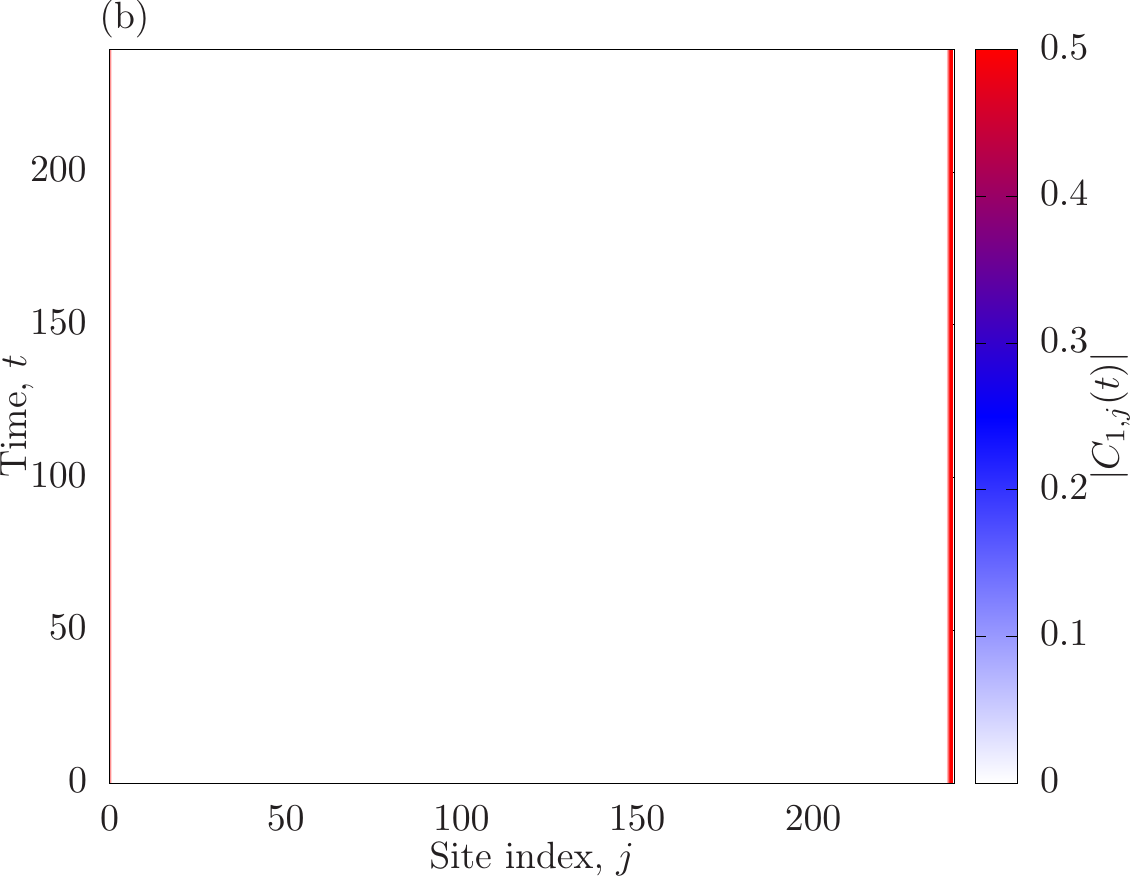}
\includegraphics[width=5.8cm]{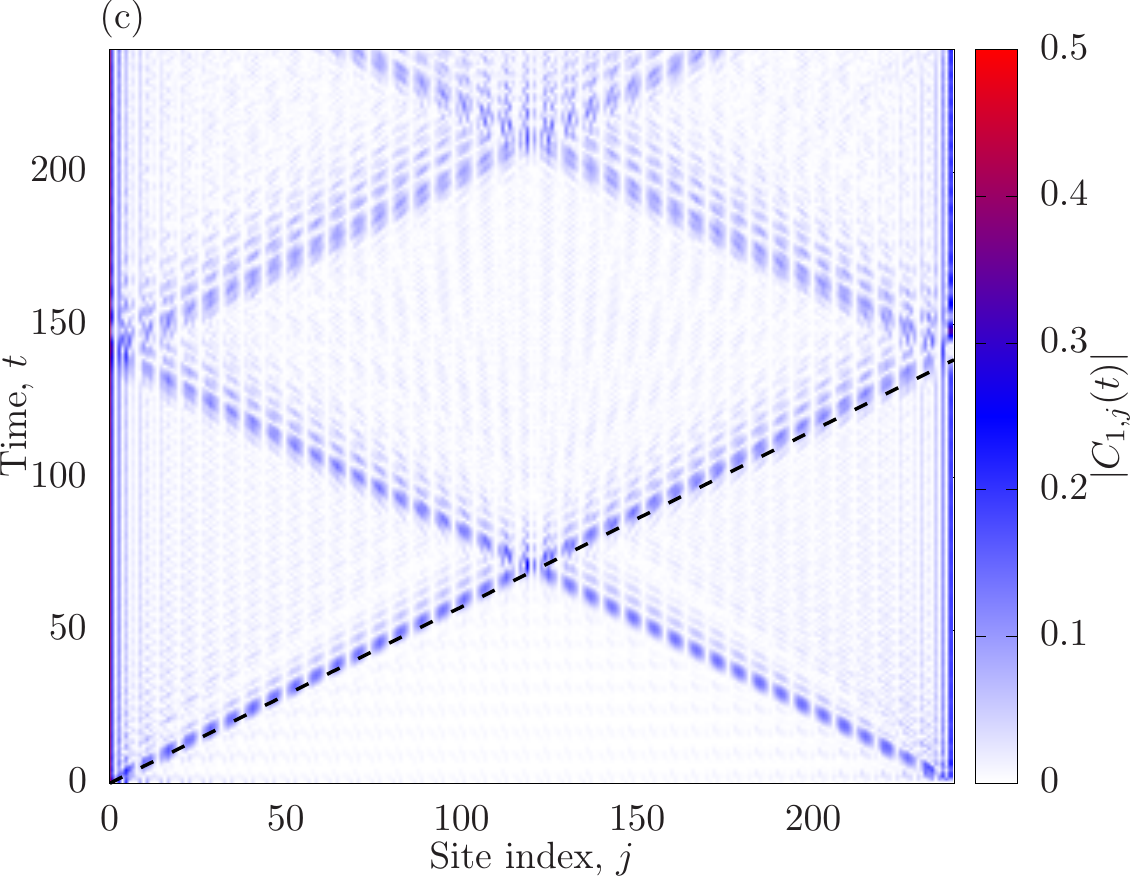}
\caption{Correlation maps $|C_{1,j}(t)|$ for dimerized states, using
  $N=240$. (a) Rainbow, (b) Frozen rainbow, (c) Island-3. The straight
  lines correspond to the theoretical predictions.}
\label{fig:corrmap_rainbow}
\end{figure*}

\beq
v_{\eff,m}=2\sin\({m\over P}\pi\),
\eeq
with $m\in \{1,\cdots,P-1\}$. Thus, it will have a single light-cone
for $P=2$ and $P=3$,  which matches with the prediction
  given in \cite{Viti.18}. However, we additionally show that for
  $P>3$ one gets more than one light-cone. 
Fig. \ref{fig:corrmap_wigner} shows that this is indeed the case,
using $N=240$. Notice that the innermost light-cone could have been
predicted just by considering the group velocity at the corresponding
filling factor, but our theoretical framework predicts all of
them. Moreover, the outermost light-cone, is not predicted by the
group velocity framework for $P>3$.

Next, let us check the validity of our results for the dimer state and
its relatives, the dimer-$q$ states. Our prediction for the dimer
state is a single light-cone, with velocity $v_\eff=2\sin(\pi/2)=2$,
which is indeed the case, as we can see in
Fig. \ref{fig:corrmap_dimer} (a). Yet, for alternating patterns of
bonds and anti-bonds, we can observe lower velocities. For the dimer-1
state we have a single velocity, $v_\eff=2\sin(\pi/4)=\sqrt{2}$, which
we can check in Fig. \ref{fig:corrmap_dimer} (b). The situation for
the dimer-2 and dimer-3 states is slightly more involved. In general,
the velocities of the dimer-$q$ states are given by

\beq
v_{\eff,p}=2\sin\( {(2p-1)\over 4q}\pi \),
\eeq
and we can see that for the dimer-2 the velocities are
$v_\eff=2\sin(\pi/8)$ and $2\sin(3\pi/8)$, as shown in
Fig. \ref{fig:corrmap_dimer} (c), while for the dimer-3, the
velocities are $v_\eff=2\sin(\pi/12)$, $2\sin(3\pi/12)$ and
$2\sin(5\pi/12)$, which are shown in Fig. \ref{fig:corrmap_dimer}
(d).

Next, let us consider the rainbow and the frozen-rainbow states. In
Fig. \ref{fig:corrmap_rainbow} (a) we see the time-evolved correlation
function $|C_{1,j}(t)|$ for the rainbow state, which is very similar
to that of the dimerized state. Indeed, our theoretical prediction is
that there will be a single light-cone with velocity
$v_\eff=2\sin(\pi/2)=2$. For the frozen rainbow our prediction is, on
the other hand, that $v_\eff=0$, which is apparent from the absence of
time-evolution in the correlation function that we can see in
Fig. \ref{fig:corrmap_rainbow} (b). Indeed, the frozen rainbow can be
proved to be an eigenstate of our Hamiltonian, Eq. \eqref{eq:ham}.

Finally, let us consider the island-3 state, which is not a valence
bond state, and is obtained as the ground state of Hamiltonian
\eqref{eq:island_ham} with $\gamma=1-10^{-3}$. Indeed, the theoretical
prediction based on the observation of the numerical form factor seen
in Fig. \ref{fig:ff} (e) is that we will obtain a single light-cone
with $v_\eff=2\sin(\pi/3)$, which can be checked in the dashed
straight line on the plot of Fig. \ref{fig:corrmap_rainbow} (c).

\begin{figure*}
  \includegraphics[width=7cm]{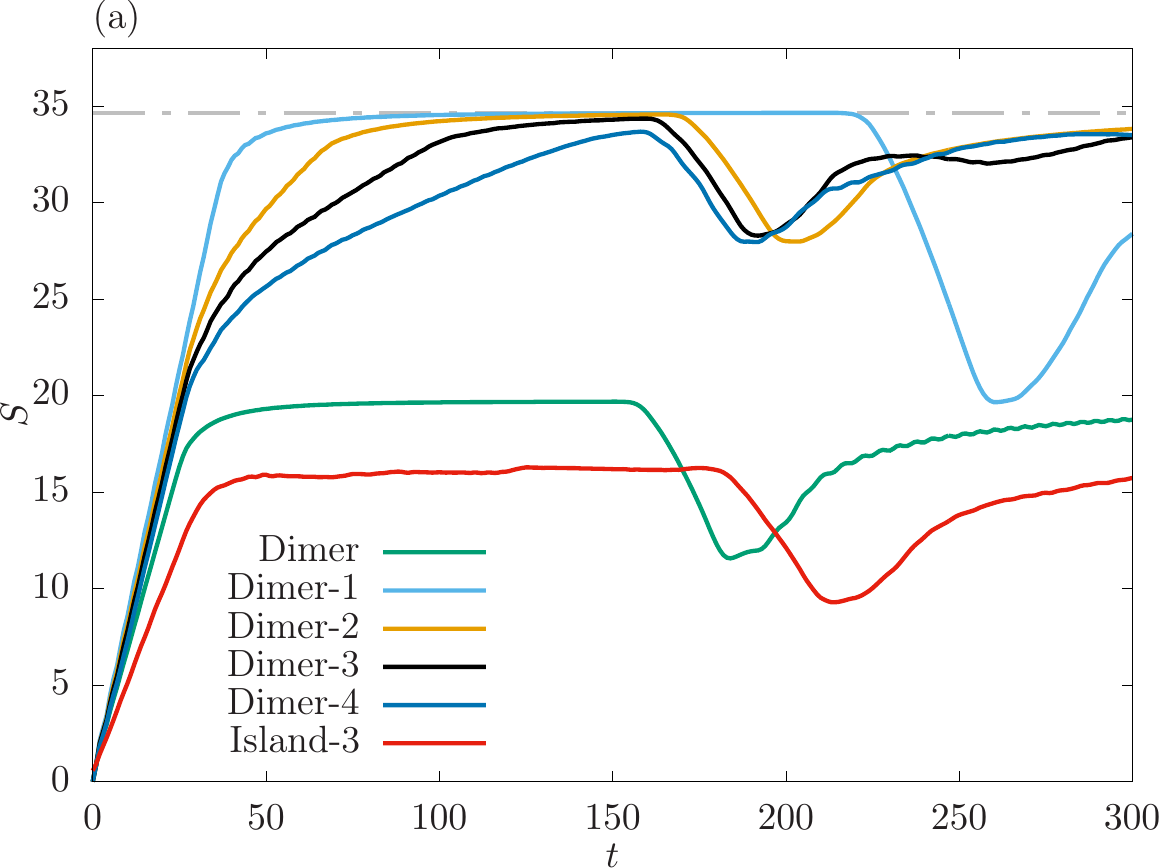}
  \includegraphics[width=7cm]{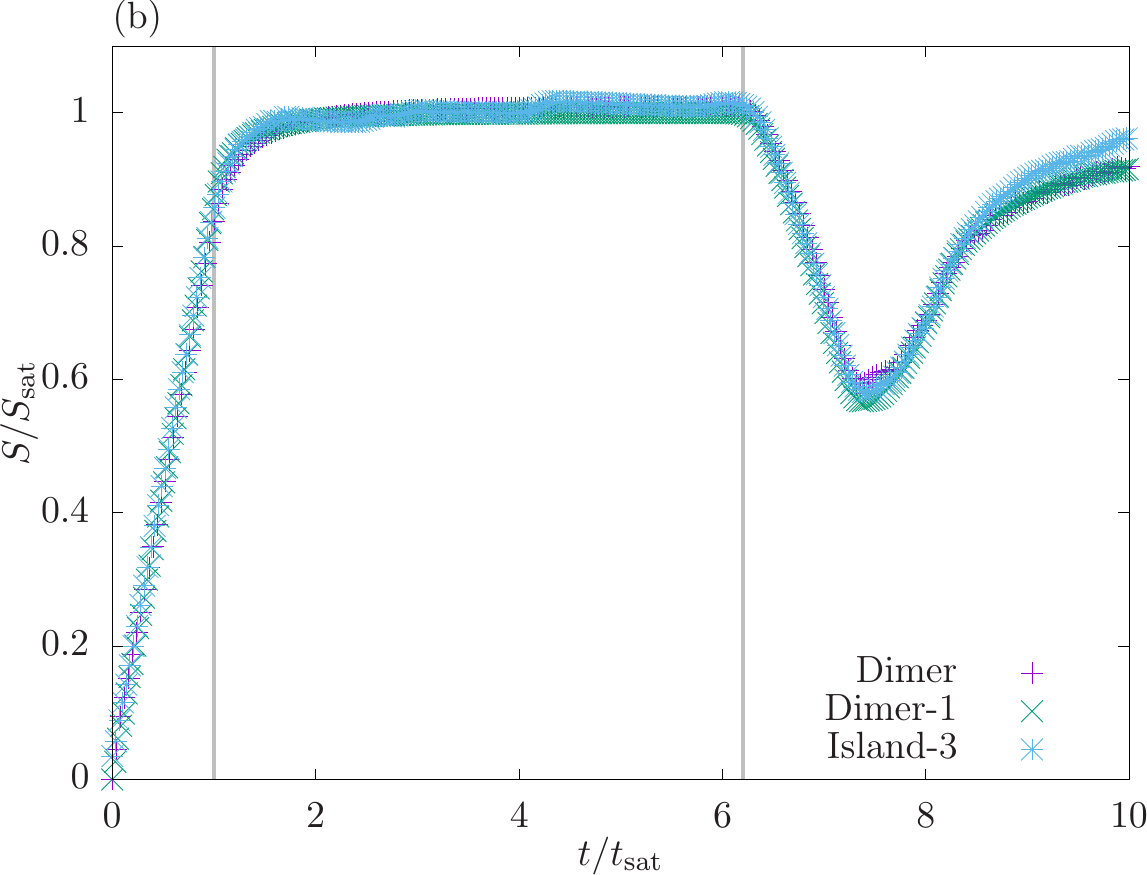}
  \caption{(a) Time-evolution of the EE of certain selected states
    under the action of Hamiltonian \eqref{eq:ham} for $N=360$ and
    $\ell=50$. Notice that in some cases, such as the dimer, dimer-1
    or island-3, the quasiparticle picture is fulfilled with a single
    velocity. Yet, for other states, such as dimer-2, dimer-3, we
    observe two different slopes, corresponding to the different types
    of quasiparticles. (b) When both the EE and time are rescaled by
    their saturation values, the data collapse for the three cases
    with a single light-cone, i.e. dimer, dimer-1 and island-3.}
  \label{fig:entropy}
\end{figure*}

\section{Entanglement growth}
\label{sec:entropy}

The previous results have an impact on our predictions for the growth
of the entanglement entropy  of a block of size $\ell$. The
quasiparticle picture devised by Cardy and Calabrese
\cite{Calabrese.05} provides the following Ansatz

\def\sat{\text{sat}}

\beq
S(\ell,t)=\begin{cases}
\sigma v t, & \text{if } t<t_\sat, \\
\sigma \ell, & \text{if } t>t_\sat.
\end{cases}
\label{eq:cc}
\eeq
where $v$ is the effective velocity of the quasiparticles, and
$\sigma$ is the entropy per site of the stationary state after the
quasiparticle wave has gone through the block. Of course, there may be
more than one type of quasiparticles, and then the total entropy can
be estimated as a sum of terms of the form
\eqref{eq:cc}. Fig. \ref{fig:entropy} (a) shows the growth of EE for
some of the states in our family, using always $N=360$ and
$\ell=50$. Indeed, we can see that in some cases the
single-quasiparticle picture is enough to predict the behavior, but
for others we observe several regimes with different slopes, which
correspond to the passage of different types of quasiparticles, with
different velocities.

Fig. \ref{fig:entropy} (a) shows the EE of a
left-most block of $\ell=50$ sites out of a system with $N=360$ as a
function of time, for several of our states. They all start in a
linear way, as predicted by the quasiparticle picture, but they grow
with different slopes. The saturation times for this first stage
differ, since they are related to the fastest light-cone velocity
present in the correlation function. In all cases, we have
$t_\sat=\ell/v_\eff$, where $v_\eff$ corresponds to the largest
effective velocity. Moreover, the saturation values for the entropy
are also very different among them, and we see that the dimer-$q$
states reach the maximal possible value, $S_\sat\approx \ell \ln(2)$,
but the dimer and the island-3 do not.

The states with several light-cones, such as the dimer-2, dimer-3 and
dimer-4, present more than one linear stage of growth, with different
slopes, related to the passage of the different types of
quasiparticles. Once the quickest ones have saturated, the slower ones
still keep entangling the block with its environment, until they also
saturate at a later time. At a time $t=(L-\ell)/v_\eff$ the quickest
particles have traveled around the whole system, and they start
meeting again inside the initial block. We start a {\em low entangling
  phase}, in which the entanglement decreases linearly, reaching a
lower value beyond which it starts growing again.

Let us consider the simplest case, that in which we obtain a single
light-cone with a single speed. Among our examples, we have the dimer
case, with $v_\eff=2$, the dimer-1 case, with
$v_\eff=2\sin(\pi/4)=\sqrt{2}$ and the island-3 case, with
$v_\eff=2\sin(\pi/3)=\sqrt{3}$. Thus, we predict that the saturation
times will be, respectively, $t_\sat=\ell/v_\eff$ in all
cases. Fig. \ref{fig:entropy} (b) shows the EE 
$S/S_\sat$ divided by the saturation value, as a function of
$t/t_\sat$, for a block $\ell=50$ from a system with $N=360$ using the
dimer, dimer-1 and island-3 states. The data for the three systems
collapse not only during the growth stage but also on the
low entangling phase, both of which are marked by vertical lines.


\section{Universal features of the correlation matrix}
\label{sec:kpz}

As we have checked both analytically and numerically, the
time-evolution of the correlation function of the discussed states
under the free-fermionic Hamiltonian presents one or several
light-cones related to the patterns present in the initial state. In
this section we will discuss the internal structure of this
time-evolved correlation function, both on the light-cone and away
from it. In the rest of this
section we will consider that time has been rescaled such that

\beq
t\to v_\eff t/2,
\eeq
and that we are considering a single contribution to the form factor,
i.e. a single value of $p$ in Eq. \eqref{eq:functional_ff}.

\begin{figure}
  \includegraphics[width=7cm]{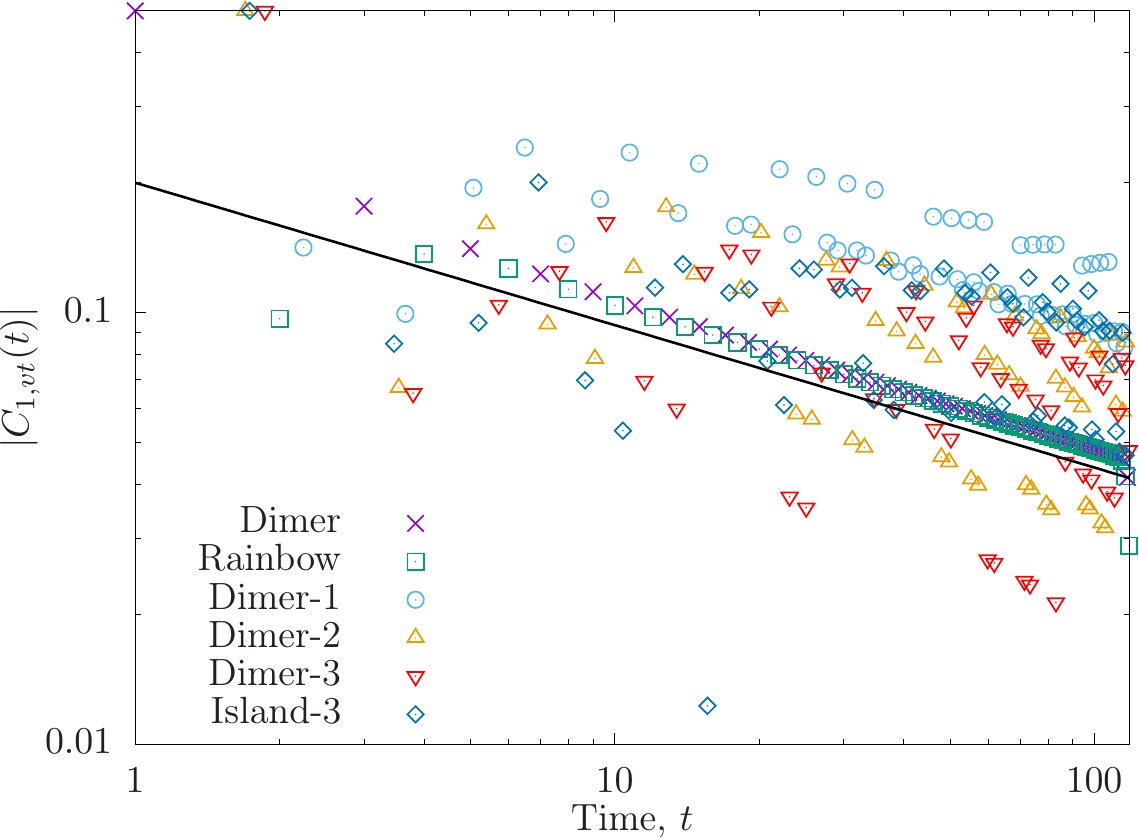}
  \caption{Decay of the correlation between site 1 and site $j=v_\eff
    t$ as a function of time for a system with $N=240$ sites, for most
    of our initial states, using always the maximal velocity $v_\eff$
    for which a light-cone appears. The continuous black straight line
    correspond to the theoretical prediction, $C\sim t^{-1/3}$.}
  \label{fig:decay}
\end{figure}

\begin{figure*}[t]
  \includegraphics[width=9cm]{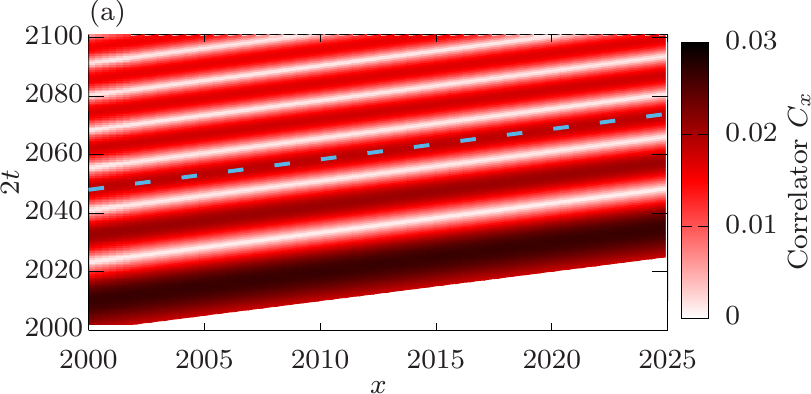}
  \includegraphics[width=7.5cm]{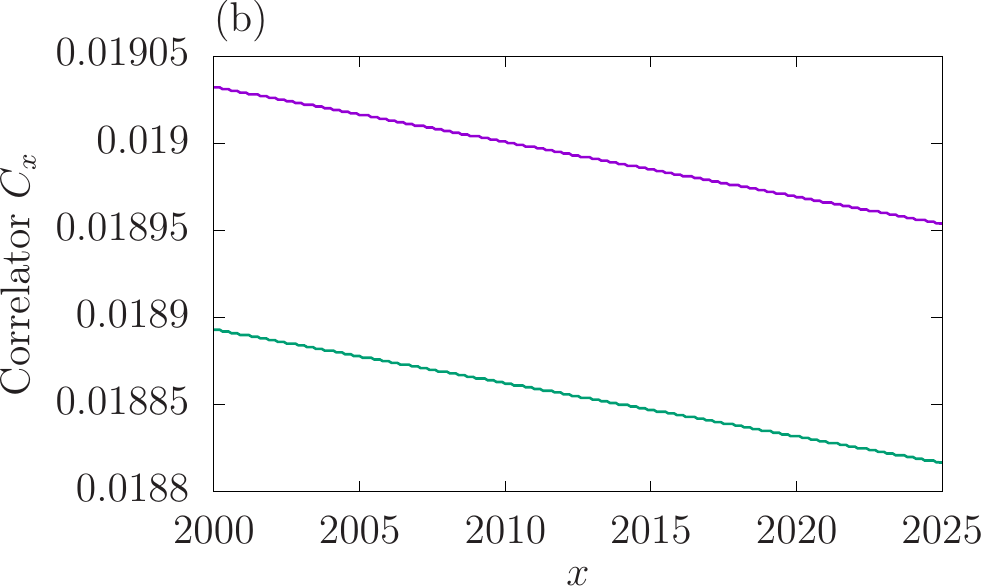}
  \caption{(a) Space-time diagram of the evolution of the correlation
    of the dimer state after a quench obtained using
    Eq. (\ref{eq:corr_bessel_new}) for the region $x\gg1$. The diagram
    shows the correlation between the left-most site with all other
    sites, $C_x=|\<\Psi(t)|c^\dagger_1c_{1+x}|\Psi(t)\>|$. Each of the
    lines approximately corresponds to a maxima (or a minima ) of
    $|C_x|$ that can be obtained by putting the argument of cosine in
    Eq. (\ref{away_corr_light_cone}) to be integer (half integer)
    multiple of $\pi$. For instance, the dashed line corresponds to a
    maxima of $|C_x|$ where the argument of cosine in
    Eq. (\ref{away_corr_light_cone}) takes the value $2\pi$. (b)
    Comparison of the values of correlator $|C_x|$ obtained using
    Eqs. (\ref{eq:corr_bessel_new}) (violet) and
    (\ref{away_corr_light_cone}) (green) along the dashed line shown
    in Fig. \ref{fig:density_corr}.}
  \label{fig:density_corr}
\end{figure*}
\subsection{Asymptotic expansion along the light cone}

Let us start with expression \eqref{eq:corr_bessel_new} for the
time-evolved correlation function in the dimer case. The asymptotic
expansions of $J_x(2t)$ in the vicinity of the light cone $x=2t$ is
given by

\beq
J_x(x) \sim
\frac{2^{1/3}}{3^{2/3} \Gamma(2/3)}
\frac{1}{x^{1/3}}, \qquad x \to\infty.
\label{eq:asympt_J_1}
\eeq
Now replacing $J_{x-1}$ and $J_{x+1}$ by $J_x$, we get

\begin{eqnarray}
C(x=2t) \approx \frac{e^{i\frac{{(j'+j+1)}\pi}{2}}} {2~3^{2/3}\Gamma(2/3)}
\frac{1}{t^{1/3}}.
\label{corr_light_cone}
\end{eqnarray}
This implies a behavior $t^{-1/3}$ for the correlator exactly on the
light cone $x=2t$. This behavior is expected in all our cases, since
the correlation matrix is a sum of terms, each one of them associated
to a light-cone, and all of them presenting a behavior similar to
\eqref{corr_light_cone}. Indeed, this is proved to be the case in
Fig. \ref{fig:decay}, where we show the decay along the light-cone of
the correlation for most of our states, always using $N=240$. For the
dimer and the rainbow the power-law decay is very clean. For all the
other states we have chosen the most intense light-cone, and the
results present oscillations which partially mask the universal
features. Yet, we can see that in all the considered cases (dimer-1,
dimer-2, dimer-3 and island states) the $t^{-1/3}$ scaling is
respected to a good approximation.

\subsection{Asymptotic expansion away from light cone}

There is another interesting asymptotic behavior that emerges
following the approximation of the Bessel functions
\cite{Abramowitz}, given by

\beq
J_\nu(\nu+z \nu^{1/3}) \sim
(2/\nu)^{ 1/3}\, \text{Ai}(-2^{1/3}z), \qquad  \nu\rightarrow \infty 
\label{eq:asympt_J_2}
\eeq
where Ai$(z')$ is the Airy function. Now one can use the asymptotic
behavior of Ai$(z')$ given by

\begin{eqnarray}
\text{Ai}(-z') \sim \pi^{-1/2} {z'}^{-1/4} \sin(\zeta + \pi/4),
 |z'| \gg 1,
 \nonumber\\
\label{eq:asympt_airy}
\end{eqnarray}
and consider $z'>0$, with $\quad \zeta = \frac{2}{3} {z'}^{3/2}$.
Plugging \eqref{eq:asympt_airy} into \eqref{eq:asympt_J_2} yields

\beq
J_\nu(\nu + z \nu^{1/3}) \sim
\frac{2^{1/4}}{\pi^{1/2} \nu^{1/3}z^{1/4}}\,
\cos\(\frac{(2z)^{3/2}}{3}-\frac{\pi}{4}\).
\label{eq:asympt_J_3}
\eeq
These expressions give the asymptotics of $J_x(2t)$ with the
identifications

\beq
x = \nu, \quad
2 t = \nu + z \nu^{1/3} \to z = \frac{2t-x}{x^{1/3}}.
\eeq
Plugging that into Eq. \eqref{eq:asympt_J_3} we get

\begin{eqnarray}
J_x(x+z x^{1/3}) &\sim &
\frac{2^{1/4}}{\pi^{1/2} x^{1/4} (2t-x)^{1/4}} \nonumber\\&&
\cos\(\frac{{ (4t-2x)}^{3/2}}{3x^{1/2}}-\frac{\pi}{4}\),
\label{eq:asympt_J_4}
\end{eqnarray}
which may lead to another prediction for the correlators within the
light cone,

\beq
C(x,t) \simeq  \frac{e^{i\frac{{(j'+j+1)}\pi}{2}}}{2^{3/4}\pi^{1/2}
  x^{1/4} (2t-x)^{1/4}} \cos\(\frac{{ (4t-2x)}^{3/2}}{3x^{1/2}}-\frac{\pi}{4}\).
\label{away_corr_light_cone}
\eeq

In Fig. \ref{fig:density_corr} (a) we plot the behavior of $C(x,t)$
obtained using Eq. (\ref{eq:corr_bessel_new}), for the region $x\gg1$
and identify that the maximum (minimum) in Fig. \ref{fig:density_corr} (a)
are obtained when the argument of cosine in
Eq. (\ref{away_corr_light_cone}) becomes integer (half integer)
multiple of $\pi$.  For instance, the dashed line in
Fig. \ref{fig:density_corr} (a) corresponds to $\frac{{
    (4t-2x)}^{3/2}}{3x^{1/2}}-\frac{\pi}{4}=2\pi$.  Hence, one can
find that along such lines, correlators again approximately behave
similar to the light-cone,

\begin{eqnarray}
C(x,t) &\simeq & \frac{e^{i\frac{{(j'+j+1)}\pi}{2}}}{2^{3/4}\pi^{1/2} x^{1/3} z^{1/4}}.
\end{eqnarray} 
An exact comparison of $|C_x|$ obtained using
Eq. \eqref{eq:corr_bessel_new} to that obtained using
Eq. \eqref{away_corr_light_cone} along this line is presented in
Fig. \ref{fig:density_corr} (b).


\section{Conclusions and further work}
\label{sec:conclusions}

We have considered the time-evolution of several quantum states on a
periodic chain with spatial patterns under the massless free fermion
Hamiltonian, finding that all of them present one or several
light-cones with different velocities, which can be read from the form
factor, i.e. the initial correlation matrix in momentm space. As we
have been able to check, in all the considered cases the form factor
is concentrated on straight lines, and the momentum shift associated
to each of them provides an effective light-cone velocity. In some
cases we were able to find a single light-cone, but with a velocity
lower than the Fermi velocity associated to the considered Hamiltonian
and filling factor, bearing some similarities to the recent
experiments in which a light beam can be seen to propagate in vacuum
with a velocity lower than $c$, due to interference effects associated
to its internal structure.

Moreover, we have found that this complex light-cone structure shows
up in the time-evolution of the entanglement entropy. In the case
of initial states which give rise to a single light-cone we were able
to collapse the EE as we rescale both time and the entropy to its
maximal saturation  value, which also differs from one state to
others. Yet, for the complex light-cone structures we observe that the
EE growth presents several linear regimes before saturation. This
behavior can be explained within the quasiparticle picture, if we
assume that there are several species of quasiparticles. We would
like to remark that the entropy production depends on the initial
state. In our case, the dimer-$q$ states reach the maximum possible
value for the entropy, $S_\sat=\ell \ln(2)$, while other states seem
to reach lower values for the saturation, implying that their
quasiparticles do not carry enough entanglement among them.

It is relevant to ask about the nature of these quasiparticles, which
is hidden inside the form factor. Indeed, it can be shown that certain
pairs of momenta are strongly entangled among themselves, and this
entanglement is preserved along the time-evolution. For example, the
frozen rainbow state entangles momenta $k$ and $-k$, and the dimer
state entangles momenta $k$ and $k\pm\pi$. An analysis of entanglement
in Fourier space would be of much help to elucidate this question
\cite{Ibanez.16}.

Beyond the existence of the light-cone and its velocity,
 we found that each straight line term in the form factor
  gives rise to a term in the correlation matrix that decays like
  $t^{-1/3}$ at large distances.  Moreover, the lost correlation
spreads away from the light-cone in a way that is also predicted by
the theory in the continuum limit.

It is relevant to ask whether these structures can be seen in
interacting systems, either integrable or non-integrable. The dynamics
of interacting systems is very different from the free theory
considered in our case, because our form factor is preserved through
the evolution. In this regard, we may conjecture that a quench to a conformally symmetric Hamiltonian will tend to give rise to a light-cone, and the spread velocity dependence on the state will also appear in those cases.


\begin{acknowledgments}
We thank Pasquale Calabrese, Erik Tonni, Vincenzo Alba for
conversations.  We also thank Jacopo Viti
 for his useful comments on our work.  We acknowledge the Spanish government for financial
support through grants PGC2018-095862-B-C21, PGC2018-094763-B-I00,
PID2019-105182GB-I00, Comunidad de Madrid grant No. S2018/TCS-4342,
SEV-2016-0597 of the ``Centro de Excelencia Severo Ochoa'' Programme
and the CSIC Research Platform on Quantum Technologies PTI-001.
\end{acknowledgments}


\appendix
\begin{widetext}
\section{Computation of the form factors}
\label{sec:appendix_ff}

In this appendix we evaluate the exact form factors for some of the
states discussed in the main text.

\subsection{Form factor for the rainbow state}
\label{sec:ff_rainbow}

Let us now compute the form factor for the rainbow state as follows.

\begin{eqnarray}
F_{k,k'} &=&\frac{1}{2N}\sum_\ell \Big(\big[(-1)^{\eta_\ell}
  e^{i(\ell k - \sigma(\ell) k')} +e^{i( k \sigma(\ell) - \ell
    k')}\Big]+\Big[e^{i(\ell k - \ell k')}+e^{i(\sigma(\ell) k -
    \sigma(\ell) k')}\Big]\Big), \nonumber\\
&=&  \frac{1}{2N}\sum_{\ell=1}^{N/2} (-1)^{\ell+N/2}  \Big[e^{i(k\ell -
    (N+1-\ell) k')} +e^{i( k (N+1-\ell)- \ell
    k')}\Big]+\frac{1}{2N}\sum_{\ell=1}^{N/2}
\Big[e^{i\ell(k-k')}+e^{i(N+1-\ell) (k-k')}\Big],
  \label{eqn:ff:rainbow2}
\end{eqnarray}
where we have used $\ell\in \mathcal{A}=\{1, 2, 3,\dots N/2\}$, and
$\sigma(\ell) \in {\bar{\cal A}}=N, N-1, \dots N+1-\ell$,
$\eta_{\ell}=N/2+\ell$. Now let us consider each term separately.

\begin{eqnarray}
&& \frac{1}{2N} \sum_{\ell=1}^{N/2} (-1)^{\ell+ N/2}  \Big[e^{i(k\ell -
      (N+1-\ell) k')}\Big] ,\nonumber\\
&=&  \frac{1}{2N} e^{-ik'(N+1)}\sum_{\ell=1}^{N/2} (-1)^{\ell+N/2}
  e^{i\ell (k+ k')}.
\end{eqnarray}
Now using the formula $\sum_{n=0}^{N/2}x^n=\frac{x^{N/2+1}-1}{x-1}$,
and thus
$\sum_{n=1}^{N/2}x^n=\frac{x^{N/2+1}-1}{x-1}-1=\frac{x^{N/2+1}-x}{x-1}$,
we get 

\begin{eqnarray}
&& \frac{(-1)^{N/2}}{2N} e^{-ik'(N+1)}\sum_{\ell=1}^{N/2} (-1)^{\ell}
  e^{i\ell (k+ k')},\nonumber\\
&=& \frac{(-1)^{N/2+1}}{2N} e^{-ik'(N+1)}\Big[ \frac{(-1)^{N/2+1}
      e^{i(N/2+1)(k+k')}+e^{i(k+k')}}{e^{i(k+k')}+1}\Big].
\label{eqn:first_term}
\end{eqnarray}
Similarly, for the second term, in Eq. (\ref{eqn:first_term}) we have
to replace $k$ by $-k'$ and we will get

\begin{eqnarray}
&&\frac{(-1)^{N/2}}{2N} e^{ik(N+1)}\sum_{\ell=1}^{N/2} (-1)^{\ell}
  \Big[e^{-i\ell(k+k')}\Big] ,\nonumber\\
&=&\frac{(-1)^{N/2+1}}{2N}  e^{ik(N+1)}\Big[ \frac{(-1)^{N/2+1}
      e^{-i(N/2+1)(k+k')}+e^{-i(k+k')}}{e^{-i(k+k')}+1}\Big],
  \nonumber\\
\end{eqnarray}
Whereas the third and the fourth term give

\begin{eqnarray}
&& \frac{1}{2N}\sum_{\ell=1}^{N/2}
  \Big[e^{i\ell(k-k')}+e^{i(N+1-\ell) (k-k')}\Big],\nonumber\\
&=&  \frac{\delta_{k,k'}}{2}.
\end{eqnarray}
Now plugging these back in Eq. (\ref{eqn:ff:rainbow2}), we get

\begin{eqnarray}
F_{k,k'}&=&\frac{(-1)^{N/2}}{2N}\sum_{\ell=1}^{N/2} (-1)^{\ell}
\Big[e^{i(k\ell - (N+1-\ell) k')} +e^{i( k (N+1-\ell)- \ell
    k')}\Big]+\frac{1}{2N}\sum_{\ell=1}^{N/2}
\Big[e^{i\ell(k-k')}+e^{i(N+1-\ell) (k-k')}\Big],\nonumber\\
&=&\frac{(-1)^{N/2+1}}{4N} \frac{
  e^{i\frac{k-k'}{2}}\Big(e^{ikN/2}+(-1)^{N/2+1} e^{-ik'N/2}\Big)^2}
{\cos(\frac{k+k'}{2})}+ \frac{\delta_{k,k'}}{2}.\nonumber\\
\end{eqnarray}


\subsection{Form factor for frozen-rainbow}
\label{sec:ff_frozen_rainbow}

The form factor for the frozen-rainbow state for $|k+k'|\neq0,
2\pi, 4\pi$ is given by

\begin{eqnarray}
F_{k,k'}&=&\frac{1}{2N}\sum_{\ell=1}^{N/2}  \Big[e^{i(k\ell -
    (N+1-\ell) k')} +e^{i( k (N+1-\ell)- \ell
    k')}\Big]+\frac{1}{2N}\sum_{\ell=1}^{N/2}
\Big[e^{i\ell(k-k')}+e^{i(N+1-\ell) (k-k')}\Big],\nonumber\\
&=&\frac{1}{2N} \Big[e^{-ik'(N+1)}
  \frac{e^{i(k+k')(N/2+1)}-e^{i(k+k')}}{e^{i(k+k')}-1} +e^{ik(N+1)}
  \frac{e^{-i(k+k')(N/2+1)}-e^{-i(k+k')}}{e^{-i(k+k')}-1} \Big]
\nonumber\\
&=&\frac{1}{2N} \Big[
  \frac{e^{ik(N/2+1)-ik'N/2}-e^{ik-ik'N}}{e^{i(k+k')}-1}
  +\frac{e^{ikN/2-ik'(N/2+1)-}-e^{ikN-ik'}}{e^{-i(k+k')}-1}\Big]+\frac{\delta_{k,k'}}{2}.
 \label{eq:ff:antirainbow1}
\end{eqnarray}
Whereas, for $|k+k'|=0, 2\pi, 4\pi$ $F_{k,k'}$ is given by

\begin{eqnarray}
F_{k,k'}&=&\frac{e^{-ik'(N+1)}+e^{ik(N+1)}}{4}+\frac{\delta_{k,k'}}{2},\nonumber\\
&=&\frac{e^{-ik'(N+1)}}{2}\delta_{|k+k'|,0}+\frac{\delta_{k,k'}}{2}.
\label{eq:ff:antirainbow2}
\end{eqnarray}

\subsection{Form factor for the Dimer-1}
We derive the form factor for dimer-1 state as follows. 
\label{sec:ff_dimer1}
\begin{eqnarray}
 F_{k,k'} &=&\frac{1}{2N}\sum_{\ell=1}^{N/2} \Big(\big[(-1)^{\eta_\ell}  e^{i(\ell k - \sigma(\ell) k')} +e^{i( k \sigma(\ell) - \ell k')}\Big]+\Big[e^{i(\ell k - \ell k')}+e^{i(\sigma(\ell) k - \sigma(\ell) k')}\Big]\Big), \nonumber\\
  &=&\frac{1}{2N}\sum_{\ell=1} ^{N/2}\Big(\big[(-1)^{\ell}  e^{i[(2\ell-1) k - 2\ell k']} +e^{i[k 2\ell - (2\ell-1) k']}\Big]+\Big[e^{i(\ell k - \ell k')}+e^{i(\sigma(\ell) k - \sigma(\ell) k')}\Big]\Big),\nonumber\\
    &=&\frac{1}{2N} \Big(e^{-ik}\sum_{\ell=1}^{N/2} (-1)^{\ell}  e^{i2\ell(k-k') } +  e^{ik'} \sum_{\ell=1}^{N/2}  (-1)^{\ell} e^{i(\ell k - \ell k')}\Big)+\frac{\delta_{|k-k'|,0}}{2},\nonumber\\
F_{k,k'} &=& \frac{e^{-ik}+e^{ik'}}{4} (\delta_{|k-k'|,\pi/2}+\delta_{|k-k'|,3\pi/2})+\frac{\delta_{|k-k'|,0}}{2}.
\label{FF:example1}
\end{eqnarray}

\subsection{Form factor for the Dimer-2 and any general Dimer-q state }
We now aim to derive the form factor of  dimer-q state  for any general $q$. For that we first present the case for $q=2$. 
\begin{eqnarray}
  F_{k,k'} &=&\frac{1}{2N}\Big( \sum_{\ell=1}^{N/4}(-1)^{\eta_\ell}  \Big[e^{i( k(4\ell-3) - k'(4\ell-2))} +e^{i( k(4\ell-1) - k'(4\ell))} \Big]+\sum_{\ell=1}^{N/4}(-1)^{\eta_\ell}  \Big[e^{i( k(4\ell-2) - k'(4\ell-3))} +e^{i( k(4\ell) - k'(4\ell-1))} \Big]\nonumber\\
  &+&\sum_{\ell=1}^{N/2}\Big[e^{i(\ell k - \ell k')}+e^{i(\sigma(\ell) k - \sigma(\ell) k')}\Big]\Big),\\
  &=&\frac{1}{2N}\Big( e^{-i(3k-2k')}\sum_{\ell=1}^{N/4}(-1)^{\eta_\ell}  e^{i( 4\ell(k-k') )} +e^{-ik}\sum_{\ell=1}^{N/4}(-1)^{\eta_\ell}  e^{i( 4\ell(k-k') )} +e^{i(3k'-2k)}\sum_{\ell=1}^{N/4}(-1)^{\eta_\ell}  e^{i( 4\ell(k-k') )} \nonumber\\&+&e^{ik'}\sum_{\ell=1}^{N/4}(-1)^{\eta_\ell}  e^{i( 4\ell(k-k') )}\Big)+\frac{\delta_{k,k'}}{2},\nonumber\\
    &=&\frac{1}{8}\Big( e^{-i(3k-2k')}(\delta_{|k-k'|,\frac{\pi}{4}}+\delta_{|k-k'|,\frac{3\pi}{4}}+\delta_{|k-k'|,\frac{5\pi}{4}}+\delta_{|k-k'|,\frac{7\pi}{4}})+e^{-ik}(\delta_{|k-k'|,\frac{\pi}{4}}+\delta_{|k-k'|,\frac{3\pi}{4}}+\delta_{|k-k'|,\frac{5\pi}{4}}+\delta_{|k-k'|,\frac{7\pi}{4}})\nonumber\\&+&e^{i(3k'-2k)}(\delta_{|k-k'|,\frac{\pi}{4}}+\delta_{|k-k'|,\frac{3\pi}{4}}+\delta_{|k-k'|,\frac{5\pi}{4}}+\delta_{|k-k'|,\frac{7\pi}{4}})+e^{ik'}(\delta_{|k-k'|,\frac{\pi}{4}}+\delta_{|k-k'|,\frac{3\pi}{4}}+\delta_{|k-k'|,\frac{5\pi}{4}}+\delta_{|k-k'|,\frac{7\pi}{4}})  \Big),\nonumber \\&+&\frac{\delta_{k,k'}}{2},\\
    &=& \frac{1}{8}\Big(\delta_{|k-k'|,\frac{\pi}{4}}+\delta_{|k-k'|,\frac{3\pi}{4}}+\delta_{|k-k'|,\frac{5\pi}{4}}+\delta_{|k-k'|,\frac{7\pi}{4}}\Big) \Big(e^{-i(3k-2k')}+e^{-ik}+e^{i(3k'-2k)}+e^{ik'}\Big)+\frac{\delta_{k,k'}}{2}.
\label{FF:example2}
\end{eqnarray}

Hence, the form factor for any general dimer-q state  is given by 
\beq
F_{k,k'} ={\delta_{k,k'}\over 2} +{1\over 4q}\(\sum_{p=1}^q
\(e^{-i((2p-1)k-2(p-1)k')}+e^{-i(2(p-1)k-(2p-1)k')}\)\)
\(\sum_{p=1}^q
\(\delta_{|k-k'|,{\pi(2p-1)\over 2q}}
  + \delta_{|k-k'|,2\pi-{\pi(2p-1)\over 2q}}\)\).
\eeq

\end{widetext}



\end{document}